\documentclass[12pt,a4paper]{article}  
\usepackage{a4wide}
\usepackage{graphicx}
\usepackage{subfigure}
\usepackage[centertags]{amsmath}
\usepackage{amssymb}
\usepackage{cite}
\usepackage{epsfig}
\usepackage{epsfig}

\newcommand{\ketAkin}[1]{|#1\rangle}

\begin{document}  
\begin{flushright}  
                     hep-th/0406208
\end{flushright}  
\vskip 2cm    

\begin{center}  
{\huge Heterotic Brane World}   
\vspace*{5mm} \vspace*{1cm}   
\end{center}  
\vspace*{5mm} \noindent  
\vskip 0.5cm  
\centerline{\bf Stefan F\"orste, Hans Peter Nilles, Patrick
  Vaudrevange, Ak\i{}n Wingerter}
\vskip 1cm
\centerline{\em Physikalisches Institut, Universit\"at Bonn}
\centerline{\em Nussallee 12, D-53115 Bonn, Germany}
\vskip2cm
  
\centerline{\bf Abstract}  
\vskip .3cm

Orbifold compactification of heterotic $\text{E}_8 \times \text{E}_8'$ string theory
is a source for promising  
grand unified model building. It can accommodate the successful aspects
of grand unification while 
avoiding problems like doublet-triplet splitting in the Higgs
sector. Many of the phenomenological  
properties of the 4-dimensional effective theory find an explanation
through the geometry of compact space  
and the location of matter and Higgs fields. These geometrical
properties can be used as a guideline   
for realistic model building.

\vskip .3cm

\newpage

\section{Introduction}

Superstring theories are candidates for a grand unified description of
all fundamental interactions. The complexity of these theories as well
as our limited  mathematical tools, however, makes it difficult
to construct explicit models for the generalization of the
(supersymmetric) standard model of particle physics. One of the major
problems is a consistent and exact description of the process of
compactification of six spatial dimensions from $d=10$ to $d=4$. The
simplest scheme of torus compactification \cite{Narain:1985jj}
 does not allow for
chiral fermions in $d=4$. More elaborate schemes such as the compactification
on 6-dimensional Calabi-Yau manifolds \cite{Candelas:en}
allow explicit
calculations only in a limited number of cases and the road to (semi)
realistic model building is still very difficult. The concept of
orbifold compactification of the heterotic string 
\cite{Dixon:jw,Dixon:1986jc}   is more
successful, as it combines the simplicity of torus compactification with
the presence of realistic gauge groups and particle spectra in $d=4$.
The simplest  
orbifolds obtained by just twisting the $d=6$ torus, however, lead
only to a limited number of models with usually too large gauge 
groups and too
many families of quarks and leptons. A breakthrough towards realistic
model building was achieved by the inclusion of background fields
such as Wilson lines \cite{Ibanez:1986tp}. 
This scheme provides a mechanism for
(further) gauge symmetry breakdown and, more importantly and
surprisingly, an efficient way to control the number of families of
quarks and leptons. It is no longer difficult to arrive at models with a
3 family structure \cite{Ibanez:1986tp} 
and gauge group like $\text{SU}(3)\times \text{SU}(2)\times \text{U}(1)$
\cite{Ibanez:1987sn}.

In these constructions the unification of gauge coupling
constants originates from
the presence of a higher dimensional grand unified (GUT) gauge group
(e.g.\ $\text{E}_8$ or $\text{SO}(32)$) and does not necessarily include a GUT group in $d=4$;
instead only $\text{SU}(3)\times \text{SU}(2)\times \text{U}(1)$ is realized as the 
unbroken gauge group in four space-time dimensions. This
scheme has the advantage of the appearance of  incomplete
representations (so-called split multiplets)
with respect to the higher dimensional gauge group.
Among other things this allowed an elegant solution of the notorious
doublet-triplet splitting problem \cite{Ibanez:1987sn}
of $\text{SU}(5)$ GUTs, where at low energy
the Higgs-multiplet as a doublet of $\text{SU}(2)$ could be realized in the
absence of its $\text{SU}(5)$ partner (a triplet of $\text{SU}(3)$). Recently, this mechanism
has been revived 
\cite{Kawamura:2000ir,Kawamura:2000ev,Altarelli:2001qj,Hall:2001pg,Kawamoto:2001wm,Hebecker:2001wq,Asaka:2001eh} 
in a pure field theory language in $d=5$
and 6. This construction of so-called orbifold GUTs, however, requires a
careful discussion of the consistency of the field theory description
and needs a number of ad hoc assumptions. To avoid such problems, it
would therefore be advisable to embed these models in the framework of
consistent higher dimensional string theories.

Orbifold compactifications of the heterotic string theory are simple
enough to allow for a number of explicit calculations relevant for the
phenomenological properties of the scheme. This includes:

\begin{itemize}
\item
the determination of Yukawa couplings with world sheet instanton
techniques 
\cite{Hamidi:1986vh,Dixon:1986qv,Lauer:1989ax,Lauer:1990tm,Burwick:1990tu,Stieberger:1992bj,Erler:1992gt} 
that incorporates a new mechanism
for a suppression of some of the couplings depending on the localization 
properties of the fields,
\item
the computation of threshold corrections for gauge coupling constants
in view of a grand unified picture
\cite{Kaplunovsky:1987rp,Dixon:1990pc,Mayr:1993kn,Mayr:1995rx},
\item
determination of selection rules for terms of the superpotential 
\cite{Font:1988tp,Font:1988mm}
necessary for the identification of the potential flat directions and
associated moduli fields.
\end{itemize}

Many of the properties of heterotic orbifolds find a nice and compelling
explanation in terms of the geometrical structure of compactified space.
The matter fields can either propagate in the full $d=10$ space (bulk
fields in untwisted sectors) or be localized at fixed ``points'' of
space-time dimension $d=4$ or $d=6$ (brane fields in twisted sectors). 
Values of Yukawa couplings, for example,
depend strongly on the relative location of quark, lepton and Higgs
fields. The number of generations of quarks and leptons reflects the
number of compactified space dimensions 
\cite{Ibanez:1986tp} and/or the number of
twisted sectors \cite{Kim:en}.             
With these intuitive rules for model
building and the potential for many explicit calculations a thorough
analysis of heterotic orbifolds seems to be a promising enterprise.
Early work in this direction has concentrated on the properties of the
$\mathbb{Z}_3$ orbifold which was used as a toy model to exhibit the
properties of 
the scheme. Some model constructions have used 
more general ${\mathbb Z} _N$ as well
as ${\mathbb Z}_N \times {\mathbb Z}_M$ orbifolds 
\cite{Font:1988mk}, but a
detailed classification of realistic 
models has not been reported so far
\cite{Bailin:pf,Bailin:1986pd,Bailin:1987xm,Bailin:1987dm,Casas:1987us,Casas:1988se,Casas:1988hb,Katsuki:1988ku,Katsuki:1989kd,Katsuki:1989ra}.
A pretty complete survey of these
attempts can be found in 
\cite{Quevedo:1996sv}, including a comprehensive list of
references. For more recent attempts at model building see
\cite{Hwang:2002hg,Kim:2003ch,Choi:2003pq,Choi:2003ag,Kim:2003hr,Kim:2004pe,Giedt:2003an,Giedt:2004wd,Choi:2004vb}.

In the present paper we shall explain that a general analysis of
heterotic orbifolds leads to many new results beyond those known in the
framework of the ${\mathbb Z}_3$ case. It reveals a web of models with
matter fields in 
the bulk ($d=10$), brane fields in $d=4$ (3-brane) or $d=6$ (5-brane in the
usual notation) as well as intersecting 5-branes in $d=4$. This results in
a multitude of models with realistic gauge groups, three families of
quarks and leptons, doublet-triplet splitting and unified coupling
constants. The picture of intersecting branes allows a connection with
the recently much discussed field theoretic orbifold GUTs 
\cite{review} and
puts them in a consistent framework, 
in case that this is possible. In
this way promising models\footnote{For related recent work 
in the framework of a heterotic ${\mathbb Z}_6$ orbifold see ref.
\cite{Kobayashi:2004ud}.}
of the bottom-up approach of field theoretic
orbifold GUTs in $d=5$ or 6 could appear as lower dimensional shadows of
the heterotic  brane world in $d=10$. With this we also hope for a better
understanding of some field theoretic results on the localization
properties of bulk fields in $d=5$ and 6 
\cite{GrootNibbelink:2002wv,GrootNibbelink:2002qp,Lee:2003mc} 
with respect to the
appearance of localized tadpoles at fixed points and fixed tori along
the lines of 
\cite{Gmeiner:2002es,GrootNibbelink:2003gb,Nibbelink:2003rc}.

The present paper is devoted to an explanation of 
the qualitative properties of
the heterotic brane world\footnote{For the discussion of brane world schemes based on Type II
strings see \cite{Uranga:nq} and references therein.}.
To
do this in a transparent way we shall use simple toy models and relegate
the attempts at realistic model building to a future publication 
\cite{future}.
In section \ref{sec:review_of_orbifolds} we shall review the rules for constructing
orbifolds with Wilson lines (the relevant technology can be found in
detail in \cite{Ibanez:1987pj}). 
Section \ref{sec:prime_orbifolds} will explain the properties of ${\mathbb
  Z}_N$ 
orbifolds (with $N$ a prime number). The toy model is based on ${\mathbb
  Z}_3$ and we
illustrate the mechanism of gauge symmetry breakdown, and the origin of the number of families along the
lines of \cite{Ibanez:1986tp,Ibanez:1987sn}. 
Section \ref{sec:zn_x_zm} is devoted to the analysis of
${\mathbb Z}_N \times {\mathbb Z}_M$
models. As our toy example we consider ${\mathbb Z}_2 \times {\mathbb
  Z}_2$ \footnote{Some work on heterotic orbifolds of the type 
${\mathbb Z}_2 \times {\mathbb Z}_2$ has recently been reported in
\cite{Donagi:2004ht}.}. It allows the 
discussion of the picture of intersecting branes and offers a multitude
of nontrivial patterns for the positions of matter and Higgs fields. For
simplicity we restrict our discussion to a model with $\text{SO}(10)$ gauge group
and 3 families of quarks and leptons. 
In section \ref{sec:exploiting_the_geometric_picture} we shall first, equipped with the brane world
picture in $d=10$, zoom in on a particular pair
of extra dimensions and interpret the model as an orbifold in $d=6$. This
allows us to exhibit the localization of matter fields at various fixed
points ($d=4$) and fixed tori (to be interpreted as bulk in the $d=6$
model). In fact, the models allow three different ways of a six
dimensional interpretation which are interrelated by the consistency
conditions of modular invariance of the underlying string theory. The
properties of a given $d=6$ model are, of course, strongly dependent on
the details of the other four compactified dimensions. Secondly, we
shall analyze the breakdown of the unified gauge group in detail. In the
heterotic theory in $d=10$ we start with the large gauge group 
$\text{E}_8\times \text{E}_8'$. 
Twists and Wilson lines reduce this to a realistic gauge group
$H$ in $d=4$ like $\text{SO}(10)$, $\text{SU}(5)$ or directly
 $\text{SU}(3)\times \text{SU}(2)\times \text{U}(1)$. Again one
finds an illuminating geometrical picture of gauge symmetry breakdown. On
the branes typically the unbroken gauge group is enhanced with respect
to $H$, and the interplay of the various branes determines the group $H$ as
the common subgroup\footnote{This is reminiscent of the discussion in
\cite{Gmeiner:2002es} 
within the framework of the so-called fixed-point equivalent models.}. 
We exhibit
this picture in detail with toy models based on the gauge group $\text{SU}(5)$
and $\text{SU}(4)\times \text{SU}(2)\times \text{SU}(2)$. Section \ref{sec:Towards_realistic_models} will be devoted to a
discussion of the potential phenomenological properties of the heterotic
brane world, including qualitative properties of gauge coupling
unification, textures for Yukawa couplings, candidates for the
appearance of an R-symmetry and the question of proton decay. In section
\ref{sec:outlook} we shall conclude with a discussion of the strategies for explicit
model building in the heterotic brane world.

\section{Review of Orbifolds}
\label{sec:review_of_orbifolds}

We will briefly review orbifold constructions, closely following \cite{Ibanez:1987pj}. In the bosonic construction, the heterotic string is described by a 10 dimensional right moving superstring, and a 26 dimensional left moving bosonic string. We will denote the 8 right moving bosonic and fermionic coordinates of the superstring in the light-cone gauge by $X_R^i$ and $\Psi_R^i$, $i=1,\ldots,8$, respectively. The left movers include 8 bosons $X_L^i$, $i=1,\ldots,8$, and another 16 bosons $X_L^I$, $I=1,\ldots,16$, which are compactified on the torus $T_{\text{E}_8\times \text{E}_8'}$ corresponding to the root lattice of $\text{E}_8\times \text{E}_8'$. (The root lattice of $\text{SO}(32)$ is also an admissible choice.)

To construct a 4 dimensional string theory, 6 dimensions are compactified on a torus $T^6$. The resulting spectrum has $\mathcal{N}=4$ supersymmetry, and is thus non-chiral. To obtain a chiral theory with $\mathcal{N}=1$ supersymmetry, one compactifies on an orbifold \cite{Dixon:jw,Dixon:1986jc}:
\begin{equation}
\mathcal{O} = T^6 \big{/}P \otimes T_{\text{E}_8\times \text{E}_8'} \big{/} G
\end{equation}
An orbifold is defined to be the quotient of a torus over a discrete set of isometries of the torus, called the {\it point group} $P$. Modular invariance requires the action of the point group to be embedded into the gauge degrees of freedom, $P \hookrightarrow G$. $G$ is in general a subgroup of the automorphisms of the $\text{E}_8\times \text{E}_8'$ Lie algebra, and is called the {\it gauge twisting group}. In the absence of outer automorphisms, the Lie algebra automorphism can be realized as a shift $X_L\mapsto X_L+V$ in the $\text{E}_8\times \text{E}_8'$ root lattice:
\begin{equation}
P \hookrightarrow G, \qquad \theta \mapsto V 
\end{equation}
An alternative description is to define an orbifold as
\begin{equation}
\mathcal{O} = \mathbb{R}^6 \big{/}S \otimes T_{\text{E}_8\times \text{E}_8'} \big{/} G,
\end{equation}
where the lattice vectors $e_\alpha$, $\alpha=1,\ldots,6$, defining the 6-torus $T^6$ have been added to the point group to form the {\it space group} $S = \left\{\left(\theta, n_\alpha e_\alpha \right) \,\big| \,\,\theta \in P, \,\,\, n_\alpha \in \mathbb{Z}\right\}$. As before, modular invariance requires the action of the space group to be embedded into the gauge degrees of freedom,
\begin{equation}
S \hookrightarrow G, \qquad \left(\theta, n_\alpha e_\alpha \right) \mapsto \left(V, n_\alpha A_\alpha \right),
\end{equation}
where the lattice vectors $e_\alpha$ are mapped to shifts $A_\alpha$ in the gauge lattice. The shifts $A_\alpha$ correspond to gauge transformations associated with the non-contractible loops given by $e_\alpha$, and are thus Wilson lines. The action of the orbifold group on all degrees of freedom is then given by
\begin{equation}
X^i \mapsto (\theta X)^i + n_\alpha e_\alpha^i, \qquad X_L^I \mapsto X_L^I + V^I + n_\alpha A_\alpha^I,
\end{equation}
where $i=3,\ldots,8,$\,\,  $I=1,\ldots,16.$ Choosing complex coordinates on the torus,
\begin{equation}
Z^1 = X^3+iX^4, \quad Z^2 = X^5+iX^6, \quad Z^3 = X^7+iX^8,
\end{equation}
the action of the point group on the space-time degrees of freedom can be neatly summarized as
\begin{equation}
Z^a \mapsto \exp\left(2\pi i v^a \right) Z^a, \quad a=1,2,3,
\label{eq:definition_of_v}
\end{equation}
where $v$ is called the {\it twist vector}.

\subsection{Consistency Conditions}

Different 4 dimensional models can be constructed depending on the choice of the compactification torus $T^6$, the point group $P$, and the embedding into the gauge degrees of freedom $P \hookrightarrow G$. There are several constraints which must be fulfilled.

{\bf The twist $\boldsymbol{\theta}$ is well-defined.} To be well-defined on the compactification torus $T^6$, $\theta$ must be an automorphism of the torus lattice, and preserve the scalar products. In other words, $\theta$ is an isometry of the torus lattice.

{\bf $\boldsymbol{\mathcal{N}}\boldsymbol{=}\boldsymbol{1}$ supersymmetry.} Acting with $\theta\in\mathbb{Z}_N$ on a spinor representation of $\text{SO(8)}$, one immediately verifies that requiring $\mathcal{N}=1$ supersymmetry amounts to demanding $\pm v^1 \pm v^2 \pm v^3 =0$ for one combination of signs ($v^i\neq 0$). In this case, one can always choose
\begin{equation}
v^1 + v^2 + v^3 =0.
\end{equation}
The generalization of these results to $\mathbb{Z}_N\times \mathbb{Z}_M$ orbifolds is given in \cite{Font:1988mk}. Requiring the twist to be well-defined on the torus, and demanding $\mathcal{N}=1$ supersymmetry, it follows that $P$ must either be $\mathbb{Z}_N$ with $N=3,4,6,7,8,12,$ or $\mathbb{Z}_N\times \mathbb{Z}_M$ with $M,N=2,3,4,6,$ and $N$ a multiple of $M$ \cite{Dixon:jw, Font:1988mk}. For $N=6,8,12,$ there are 2 different choices for the point group $P$. The lattices on which $P$ acts as an isometry are the root lattices of semi-simple Lie algebras of rank 6. In some cases, there is more than one choice of lattice for a given set of symmetries $P$. (In the $\mathbb{Z}_2\times\mathbb{Z}_2$ case, the choice of the lattice in each complex dimension is arbitrary, and the complex directions are orthogonal.)

{\bf The Embedding $\boldsymbol{P} \boldsymbol{\hookrightarrow} \boldsymbol{G}$ is a group homomorphism.} $\theta\in \mathbb{Z}_N$ implies $\theta^N=1\!\!1,$ which in turn implies that its embedding into the gauge degrees of freedom as a shift is the identity, i.e.
\begin{equation}
N\,V \in T_{\text{E}_8\times \text{E}_8'}, \quad N\,A_\alpha \in T_{\text{E}_8\times \text{E}_8'}.
\end{equation}

{\bf Modular invariance.} For the orbifold partition function to be modular invariant, following conditions on the twist, gauge shift, and Wilson lines need to be fulfilled \cite{Ibanez:1987pj}:
\begin{gather}
N\left(V^2-v^2 \right) = 0 \text{ mod } 2\notag\\
N\,V \cdot A_\alpha = 0 \text{ mod } 1\notag\\
N\,A_\alpha \cdot A_\beta = 0 \text{ mod } 1, \quad \alpha \neq \beta\notag\\
N\,A_\alpha^2 = 0 \text{ mod } 2
\label{eq:modular_invariance}
\end{gather}
These conditions can be rewritten in a more concise form as
\begin{equation}
N\left[\left(mV + n_\alpha A_\alpha \right)^2 - m^2v^2 \right] = 0 \text{ mod } 2, \quad m=0,1.
\end{equation}
Modular invariance automatically guarantees the anomaly freedom of orbifold models.

For $\mathbb{Z}_N\times\mathbb{Z}_M$ orbifolds, the above conditions for modular invariance are generalized in a straightforward way. Let $v_1$, $v_2$ denote the twist vectors of $\mathbb{Z}_N\times\mathbb{Z}_M$, and $V_1$, $V_2$ the corresponding gauge shifts. Then, the first equation in eq.\ (\ref{eq:modular_invariance}) generalizes to
\begin{gather}
N' \left[\left(kV_1+\ell V_2 \right)^2 - \left(kv_1+\ell v_2 \right)^2 \right] = 0 \text{ mod } 2,\notag\\
N' \text{ order of } kv_1+\ell v_2,\notag\\
k=0,\ldots,N-1, \quad \ell=0,\ldots,M-1.
\label{eq:mod_inv_zN_cross_zM}
\end{gather}
For the Wilson lines, the conditions in eq.\ (\ref{eq:modular_invariance}) are the same, except that they need to be fulfilled for both $V_1$, and $V_2$.

\subsection{The Spectrum}
\label{sec:spectrum}
On an orbifold, there are 2 types of strings, twisted and untwisted closed strings. An {\it untwisted string} is closed on the torus even before identifying points by the action of the twist:
\begin{equation}
X^i(\sigma + 2\pi) = X^i(\sigma) + n_\alpha e_\alpha^i, \quad i=3,\ldots,8
\end{equation}
A {\it twisted string} is closed on the torus only upon imposing the point group symmetry:
\begin{equation}
X^i(\sigma + 2\pi) = \left( \theta X(\sigma )\right)^i + n_\alpha e_\alpha^i, \quad i=3,\ldots,8
\end{equation}
From the boundary conditions, it follows that the twisted strings are localized at the points which are left fixed under the action of some element $(\theta_i,n_\alpha e_\alpha)$ of the space group $S$. These points are called the {\it fixed points} of the orbifold.  We will call the element $g\equiv (\theta_i,n_\alpha e_\alpha)$ which corresponds to some given fixed point the {\it constructing element}, and denote the states which are localized at this fixed point by $\mathcal{H}_g$.

Since the strings propagate on the orbifold, we must project onto $S\otimes G$ invariant states. We will consider the twisted and untwisted sectors separately.

{\bf Untwisted sector.} The states in the untwisted sector are those of the heterotic string compactified on a torus, where states which are not invariant under $S\otimes G$ have been projected out. The level matching condition for the massless states is given by 
\begin{equation}
\frac{1}{2}q^2 - \frac{1}{2} = \frac{1}{4}m_R^2 = \frac{1}{4}m_L^2 = \frac{1}{2} p^2 + \tilde{N} - 1 = 0,
\label{eq:masslessness_formula_untwisted}
\end{equation}
where $q$ denotes the $\text{SO}(8)$ weight vector of the right mover ground state, e.g.\ $\ketAkin{\frac{1}{2}\,\,\frac{1}{2}\,\,\frac{1}{2}\,\,\frac{1}{2}}$ or $\ketAkin{1\,\,0\,\,0\,\,0}$. Under the action of the point group, the right and left mover states will transform as $\exp(2\pi i q\cdot v)\ketAkin{q}_R$, and  $\exp(2\pi i p\cdot V)\ketAkin{p}_L$, respectively\footnote{When we take the scalar product $q\cdot v$, we shall mean $q\cdot \tilde{v}$ with $\tilde{v}=(0,v)$.}. Only states for which the product of these eigenvalues is 1 will survive the projection.
The {\it gauge bosons} are formed by combining right movers which do not transform under the action of the point group with left movers satisfying
\begin{equation}
p\cdot V = 0 \text{ mod } 1, \qquad p\cdot A_\alpha = 0 \text{ mod } 1,
\label{eq:unbroken_gauge_group_condition}
\end{equation}
giving the {\it unbroken gauge group.} Right movers which transform non-trivially combine with left movers for which
\begin{equation}
p\cdot V = k/N \text{ mod } 1, \quad k=1\ldots,N-1, \qquad p\cdot A_\alpha = 0 \text{ mod } 1,
\end{equation}
to give the {\it charged matter}. The states which include excitations for the left movers give uncharged gauge bosons (Cartan generators), the supergravity multiplet, and some number of singlets.

{\bf Twisted sectors.} Without loss of generality, let us focus on the states corresponding to the constructing element $g\equiv (\theta_i, n_\alpha e_\alpha)$. The twist acts as a shift $p \mapsto p+V_i + n_\alpha A_\alpha$ on the momentum lattice, and as $q\mapsto q+v_i$ on the right mover ground state. In addition, the number operator $\tilde{N}$ is moded. The zero point energy of the right and left movers is changed by \cite{Dixon:1986jc}
\begin{equation}
\delta c = \frac{1}{2} \sum_k \eta^k (1-\eta^k),
\label{eq:zero_point_energy_shift}
\end{equation}
where $\eta^k = v_i^k \text{ mod } 1$ so that $0 \leq \eta^k < 1$. The level matching condition for the massless states then reads
\begin{equation}
\frac{1}{2}(q+v_i)^2 - \frac{1}{2} + \delta c= \frac{1}{4}m_R^2 = \frac{1}{4}m_L^2 = \frac{1}{2} (p+V_i+n_\alpha A_\alpha)^2 + \tilde{N} - 1 + \delta c = 0.
\label{eq:mass_formula_twisted_sector}
\end{equation}
As compared to the untwisted sector, the projection conditions in the twisted sectors are slightly more complicated. Consider the subset $Z_g$ of the space group $S$ which commutes with the constructing element $g$. Acting with $Z_g$ on the orbifold, the Hilbert space $\mathcal{H}_g$ is mapped into itself. $Z_g$ should act as the identity on $\mathcal{H}_g$, thus all elements which are not invariant under $h\in Z_g$ are projected out.

If $h\in S$ does not commute with $g$, the action of $h$ changes the boundary conditions of the states in $\mathcal{H}_g$, and states in $\mathcal{H}_g$ will be mapped to states in $\mathcal{H}_{hgh^{-1}}$. To form invariant states, one starts with some state in $\mathcal{H}_g$ and considers its image in $\mathcal{H}_{hgh^{-1}}$ for all $h\in S$. In each Hilbert space, we project onto its $Z_{hgh^{-1}}$ invariant subspace. The sum of these states is then invariant under the action of the space group $S$.

%%%%%%%%%%%%%%%%%%%%%%%%%%%% Z_N orbifolds for Prime N %%%%%%%%%%%%%%%%%%%%%%%%

\section{${\mathbb Z}_N$ Orbifolds for Prime $N$}
\label{sec:prime_orbifolds}

We illustrate the discussion of the previous section by considering $\mathbb{Z}_N$ orbifolds with prime $N$, taking the $\mathbb{Z}_3$ orbifold as the paradigm.

\begin{center}
\begin{figure}[h!]
\centerline{\epsfig{figure=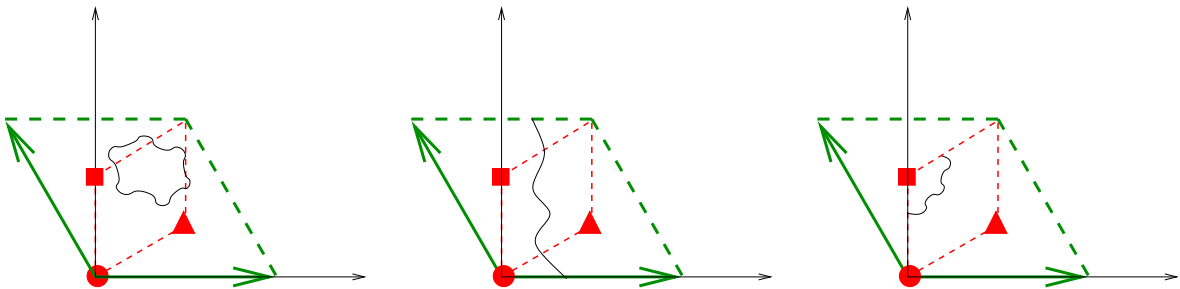,scale=.6}}
\caption{${\mathbb Z}_3$ orbifold. The circle, triangle, and square denote the fixed points.}
\label{fig:z3_orbifold}
\end{figure}
\end{center}

The lattice defining the 6-torus is the $\text{SU}(3)^3$ root lattice as shown in figure \ref{fig:z3_orbifold}. The point group $\mathbb{Z}_3$ is generated by $\theta$ which acts as a simultaneous rotation of 120${}^\circ$ in the three 2-tori, and in the notation of eq.\ (\ref{eq:definition_of_v}), this corresponds to the twist vector
\begin{equation}
v = \frac{1}{3}\left(1,\,\,\,1,\,\,-2 \right).
\end{equation}
The action of $\theta$ leaves $27$ fixed points. The twisted sector corresponding to the action of $\theta^2$ gives the anti-particles of the aforementioned sector, so we will not consider it separately.
In figure \ref{fig:z3_orbifold}, the strings in the first and the second torus are already closed on the torus (untwisted sector states), whereas in the third torus, the state only closes upon imposing the symmetry generated by the 120${}^\circ$ rotations (twisted sector state).

Let us first consider the untwisted sector. The action of the orbifold twist is accompanied by an action in the gauge degrees of freedom realized as a shift. We choose the {\it standard embedding}
\begin{equation}
V = \frac{1}{3}\left(1,\,\,\,1,\,\,-2, \,\,\,\,0^5\right)\left(0^8\right),
\label{eq:standard_embedding_z3}
\end{equation}
where the first 3 components of the gauge shift\footnote{Zero to the power of $n$ is short for writing
  $n$ zeros.} are equal to the components of the twist vector $v$. With this choice, the modular invariance condition eq.\ (\ref{eq:modular_invariance}) is automatically satisfied, and the anomaly freedom of our model is guaranteed. From the 240 states in the first $\text{E}_8$, only 78 = 72 + 6 survive the projection condition eq.\ (\ref{eq:unbroken_gauge_group_condition}), and yield the charged gauge bosons of $\text{E}_6\times \text{SU}(3)$, whereas the second $\text{E}_8'$ is left untouched. 

The right mover ground state will decompose as $\boldsymbol{8} \rightarrow \boldsymbol{3} + \bar{\boldsymbol{3}} + \boldsymbol{1} + \boldsymbol{1}$ under $\text{SU}(3) \subset \text{SO}(8)$, i.e.\ there are 3 right mover states transforming as $\ketAkin{q}_R\mapsto \exp\left(2\pi i \cdot \frac{1}{3}\right) \ketAkin{q}_R$ which will combine with left movers satisfying 
\begin{equation}
p\cdot V = \frac{2}{3} \text{ mod } 1
\end{equation}
to give the charged matter representations $3\times (\boldsymbol{27},\boldsymbol{3})$. From the untwisted sector, we thus get 9 families of quarks and leptons.

Let us now discuss the twisted sector, and focus on the fixed point $(\bullet, \bullet, \bullet)$ in figure \ref{fig:z3_orbifold}. The shift in the zero point energy as given by eq.\ (\ref{eq:zero_point_energy_shift}) is $\delta c = 1/3$, and the level matching condition for the massless states reads
\begin{equation}
\frac{1}{2}(q+v)- \frac{1}{6} = \frac{1}{4}m_R^2 = \frac{1}{4}m_L^2 = \frac{1}{2} (p+V)^2 + \tilde{N} - \frac{2}{3} = 0.
\label{eq:masslessness_formula_twisted_sector_without_wilsonlines}
\end{equation}
The twisted right moving ground state $\ketAkin{q+v}_R$ is a singlet under $\theta$. (Note that $q+v$ must be shifted by a $\text{SO}(8)$ root vector to fulfill the level matching condition). For $\tilde{N}=0$, there are 27 elements $p+V$ satisfying $(p+V)^2 = 4/3$. These left movers transform as
\begin{equation}
\ketAkin{p+V}_L\mapsto \exp\left(2\pi i(p+V)\cdot V\right)\ketAkin{p+V}_L = \exp\left(2\pi i \cdot 1\right)\ketAkin{p+V}_L,
\label{eq:transformation_of_left_movers_in_twisted_sector}
\end{equation}
and are invariant. They combine with the right mover to give the representation $(\boldsymbol{27},\boldsymbol{1})$. For $\tilde{N}=1/3$, there are 3 elements $p+V$ satisfying $(p+V)^2 = 2/3$. These left movers transform as
\begin{equation}
\ketAkin{p+V}_L\mapsto \exp\left(2\pi i(p+V)\cdot V\right)\ketAkin{p+V}_L = \exp\left(2\pi i \cdot \frac{2}{3}\right)\ketAkin{p+V}_L,
\end{equation}
whereas the oscillators (one for each complex dimension) transform as
\begin{equation}
\tilde{\alpha}^i \mapsto \exp\left(2\pi i \cdot \frac{1}{3}\right) \tilde{\alpha}^i, \quad i=1,2,3,
\end{equation}
so that the states $\ketAkin{q}_R\otimes \tilde{\alpha}^i\ketAkin{p}_L$ are invariant, and give three copies of the representation $(\boldsymbol{1},\bar{\boldsymbol{3}})$. Taking into account that there are 27 fixed points, the matter content of our orbifold model is
\begin{equation}
3 \times (\boldsymbol{27},\boldsymbol{3}), \quad 27 \times (\boldsymbol{27},\boldsymbol{1}), \quad 27\times 3\times(\boldsymbol{1},\bar{\boldsymbol{3}}).
\end{equation}
Thus, in the case of the standard embedding, we have 36 generations of quarks and leptons. All non-trivial embeddings of the point group into the gauge degrees of freedom have been classified \cite{Dixon:1986jc}. For each model, we have listed the shift vector $V$, the resulting unbroken gauge group, and the number of generations in table \ref{fig:z3_classification}.
\begin{table}[h!]
\renewcommand{\baselinestretch}{2}
\normalsize
\begin{center}
\begin{tabular}{|c|l|l|c|}
\hline
Case	& Shift $V$	& 	Gauge Group &  Gen.\\
\hline
\hline
1& $\left(\frac{1}{3}, \frac{1}{3}, \frac{2}{3}, 0^5 \right)\left(0^8 \right)$ & $\text{E}_6 \times \text{SU}(3) \times \text{E}_8'$  &  36\\
\hline
2& $\left(\frac{1}{3}, \frac{1}{3}, \frac{2}{3}, 0^5 \right)\left(\frac{1}{3}, \frac{1}{3}, \frac{2}{3}, 0^5 \right)$ & $\text{E}_6 \times \text{SU}(3) \times \text{E}_6' \times \text{SU}(3)'$  &  9\\
\hline
3& $\left(\frac{1}{3}, \frac{1}{3}, 0^6 \right)\left(\frac{2}{3}, 0^7 \right)$ & $\text{E}_7 \times \text{U}(1) \times \text{SO}(14)' \times \text{U}(1)'$  &  0\\
\hline
4& $\left(\frac{1}{3}, \frac{1}{3}, \frac{1}{3}, \frac{1}{3}, \frac{2}{3}, 0^3 \right)\left(\frac{2}{3}, 0^7 \right)$ & $\text{SU}(9)\times \text{SO}(14)' \times \text{U}(1)'$  &  9\\
\hline
\end{tabular}
\end{center}
\renewcommand{\baselinestretch}{1}
\caption{Inequivalent $\mathbb{Z}_3$ orbifold models without Wilson lines.}
\label{fig:z3_classification}
\end{table}

Note that the proliferation of the number of generations is due to the fact that the physics at each fixed point is the same. This dramatically changes in the presence of Wilson lines \cite{Ibanez:1986tp}. We will illustrate the lifting of the degeneracy at the fixed points using a specific example. Choose the standard embedding, and the Wilson lines
\begin{equation}
A_1 = A_2 = \left(0^6,\,\, \frac{1}{3} ,\,\,\frac{1}{3} \right)\left(\frac{2}{3} ,\,\,0^7 \right).
\end{equation}
Applying the projection conditions eq.\ (\ref{eq:unbroken_gauge_group_condition}), we find that the surviving gauge symmetry is
\begin{equation}
\text{SU}(6)\times \text{SU}(3)\times \text{U}(1)\times \text{SO}(14)'\times \text{U}(1)'.
\end{equation}
From the untwisted sector, we obtain the charged matter representations $3\times (\boldsymbol{15},\boldsymbol{3})$. (We will only indicate the representations under the first 2 factors of the symmetry group.) Let us discuss the twisted sector in greater detail.

\begin{figure}[h!]
\begin{center}
\centerline{\epsfig{figure=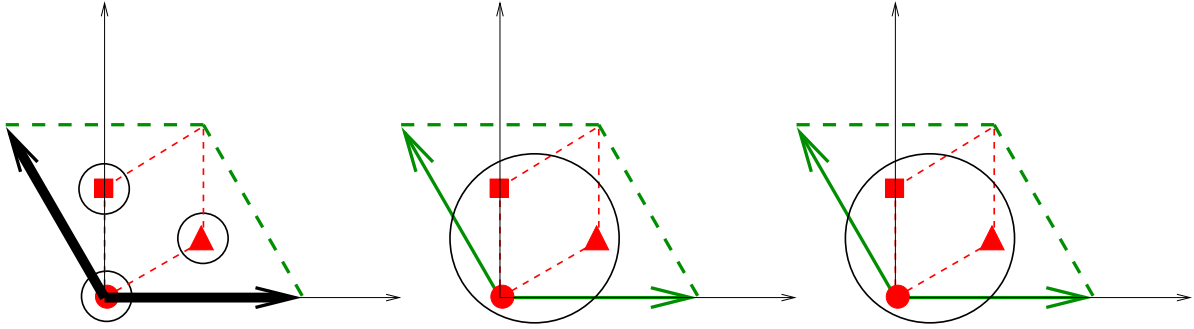}}
\caption{${\mathbb Z}_3$ orbifold with non-vanishing Wilson lines $A_1, A_2$. The circles around the fixed points indicate that the degeneracy in the first torus is lifted.}
\label{fig:z3_orbifold_with_wilson_lines}
\end{center}
\end{figure}

Consider the fixed points $(\bullet, \bullet, \bullet)$, $(\blacktriangle, \bullet, \bullet)$, and $({\scriptstyle \blacksquare}, \bullet, \bullet)$ as depicted in figure \ref{fig:z3_orbifold_with_wilson_lines}. The fixed point $(\bullet, \bullet, \bullet)$ is left invariant under the action of $\theta$ alone, i.e.\ the constructing element is $(\theta,0)$. The level matching condition for massless states living at this fixed point will be the same as eq.\ (\ref{eq:masslessness_formula_twisted_sector_without_wilsonlines}). The states do not feel the presence of the Wilson lines.
The fixed point $(\blacktriangle, \bullet, \bullet)$, however, is only invariant under the action of $\theta$ accompanied by the lattice shift $e_1$, and the constructing element is $(\theta,e_1)$. The immediate consequence is that the level matching condition for the massless states changes to
\begin{equation}
\frac{1}{2}(q+v)- \frac{1}{6} = \frac{1}{4}m_R^2 = \frac{1}{4}m_L^2 = \frac{1}{2} (p+V+A_1)^2 + \tilde{N} - \frac{2}{3} = 0.
\end{equation}
Clearly, it is more difficult to fulfill the new relation, and the $\boldsymbol{27}$ of $\text{E}_6$ will preferentially decompose into small representations under the new gauge group. The level matching condition can only be satisfied for $\tilde{N}=0$, and these states also survive the projection condition analogous to eq.\ (\ref{eq:transformation_of_left_movers_in_twisted_sector}) (where we have to substitute $V\rightarrow V+A_1$) to form the representations $(\boldsymbol{1}, \bar{\boldsymbol{3}}) + (\bar{\boldsymbol{6}}, \boldsymbol{1})$. As there are no Wilson lines in the second and third torus, the spectrum at the fixed point $(\blacktriangle, \bullet, \bullet)$ will still be 9-fold degenerate. All fixed points with $\blacktriangle$ as the first entry and an arbitrary one in the last two entries will have the same matter content. Analogous considerations also apply in the case of the fixed point $({\scriptstyle \blacksquare}, \bullet, \bullet)$.

To summarize, the matter content of the model is (omitting the antiparticles)
\begin{center}
\renewcommand{\baselinestretch}{1.5}
\normalsize
\begin{tabular}{ll}
\text{Untwisted} & $3\times (\boldsymbol{15}, \boldsymbol{3})$,\\
$(\bullet, \cdot, \cdot)$ & $9\times (\boldsymbol{15}, \boldsymbol{1})$, $18\times (\bar{\boldsymbol{6}}, \boldsymbol{1})$, $27\times (\boldsymbol{1}, \bar{\boldsymbol{3}})$,\\
$(\blacktriangle, \cdot, \cdot)$ & $9\times (\boldsymbol{1}, \bar{\boldsymbol{3}})$, $9\times (\bar{\boldsymbol{6}}, \boldsymbol{1})$,\\
$({\scriptstyle \blacksquare}, \cdot, \cdot)$ & $9\times (\boldsymbol{1}, \bar{\boldsymbol{3}})$, $9\times (\bar{\boldsymbol{6}}, \boldsymbol{1})$.
\end{tabular}
\end{center}
From the untwisted sector, we obtain 9 families, which also have
$\text{SU}(3)$ quantum numbers, and from $(\bullet, \cdot, \cdot)$, we
have another 9 families which are $\text{SU}(3)$ singlets. The total
number of 18 families is to be compared to the 36 families in the case
of no Wilson lines. Using more Wilson lines, models  with 3 families
of quarks and leptons \cite{Ibanez:1986tp} and with standard model gauge group
$\text{SU}(3)\times\text{SU}(2)\times\text{U}(1)^n$ can be constructed
\cite{Ibanez:1987sn}.

\clearpage

\section{${\mathbb Z}_N\times {\mathbb Z}_M$ Orbifolds}
\label{sec:zn_x_zm}

In the previous section, we discussed ${\mathbb Z}_N$ orbifolds with $N$ being 
a prime number. Some additional structure arises when $N$ is not prime, or for 
${\mathbb Z}_N\times {\mathbb Z}_M$ orbifolds. These theories have ${\cal N} = 2$ 
subsectors, because the point group contains elements for which one entry of the 
corresponding twist vector $v$ vanishes. Actually, the fixed points under these 
elements are fixed tori. As the simplest example, we discuss a ${\mathbb Z}_2\times 
{\mathbb Z}_2$ orbifold\footnote{For a recent discussion of ${\mathbb Z}_2\times {\mathbb Z}_2$
twists in the fermionic formulation of heterotic string theory
see \cite{Faraggi:2004rq}.}.

\begin{figure}
\centerline{\epsfig{figure=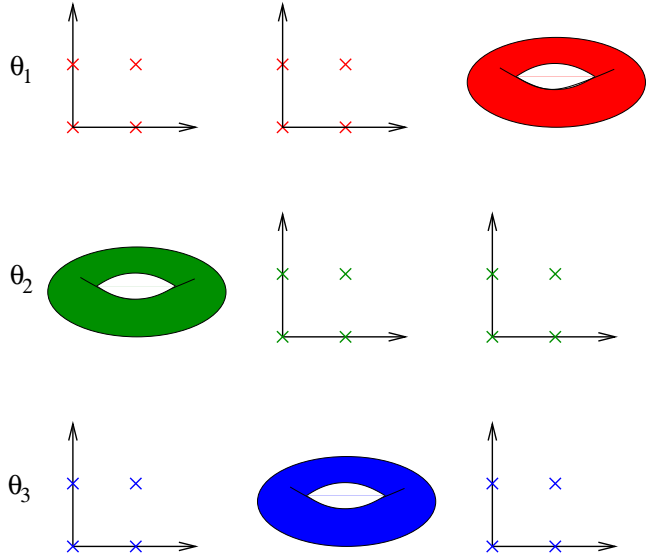,angle=270,scale=.4}}
\caption{Action of the twists in the case of the ${\mathbb Z}_2\times {\mathbb Z}_2$ orbifold. The crosses indicate the fixed points.}
\label{fig:actionoftwist}
\end{figure}

\subsection{${\mathbb Z}_2\times {\mathbb Z}_2$ Orbifolds}
The ${\mathbb Z}_2\times {\mathbb Z}_2$ point group consists of four elements: 
$1\!\!1$, $\theta_1$, $\theta_2$, and $\theta_3 = \theta_1 \theta_2$. Their action 
is given by rotations in three complex planes (see figure \ref{fig:actionoftwist}): 
$v_1  = \left( \frac{1}{2}, -\frac{1}{2}, 0\right)$, $v_2 = \left( 0, \frac{1}{2}, 
-\frac{1}{2}\right)$ and $v_3 = v_1 + v_2$. Each of these twists acts only in 
two of the three complex planes, creating $4 \cdot 4 = 16$ fixed points. 
Therefore, the strings of the twisted sectors are only fixed in four of the six compact
dimensions, still free to move in two of them. Thus, the 16 fixed points of every
twisted sector are in fact 16 fixed tori. Altogether, the {${\mathbb Z}_2\times 
{\mathbb Z}_2$ orbifold has 48 fixed tori. Counting also the $3 + 1$ non-compact 
dimensions, these are actually $5 + 1$ dimensional fixed planes, i.e.\ 5-branes. 
Branes belonging to different twists are mutually orthogonal, and intersect in 
$4d$ Minkowski space. A picture showing one brane from each twisted sector 
is given in figure \ref{fig:threebranes}. The 16 5-branes from the same twisted 
sector are parallel to each other.

\begin{figure}
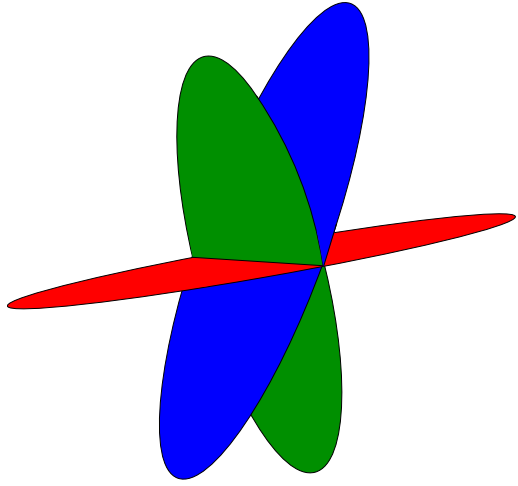
  
\center\input intersecting_branes.pstex_t
\caption{Intersecting brane picture: The picture shows one brane of each twisted 
sector. The intersection is $4d$ Minkowski space. The branes are mutually orthogonal 
in the six compact dimensions.}
\label{fig:threebranes}
\end{figure}
Any of the twists break $\mathcal{N} = 4$ to $\mathcal{N} = 2$ supersymmetry. The combination 
of all twists leaves $\mathcal{N} = 1$ supersymmetry unbroken.

The twists must be embedded into the gauge degrees of freedom such that
\begin{eqnarray}
2 \left[ (k V_1 + l V_2)^2 - (k v_1 + l v_2)^2\right] & = & 0 \textmd{ mod } 2, \qquad k, l = 1, 2
\end{eqnarray}
holds, in order to ensure modular invariance (eq.\ (\ref{eq:mod_inv_zN_cross_zM})). The easiest way to fulfill 
this condition is the standard embedding 
\begin{eqnarray}
V_1 & = & \left( \frac{1}{2}, -\frac{1}{2}, 0^6\right) \left(0^8\right), \nonumber\\
V_2 & = & \left( 0, \frac{1}{2}, -\frac{1}{2}, 0^5\right) \left(0^8\right), \nonumber\\
V_3 & = & V_1 + V_2. \nonumber
\end{eqnarray}
The \textbf{untwisted sector} is given by the spectrum of the heterotic string, projected onto 
$S\otimes G$ invariant states. These states are now sorted with respect to their eigenvalues. The 
eigenvalues for right and left movers are given by
\begin{eqnarray}
\exp(2\pi i q \cdot v_i) \ketAkin{q}_R & \quad \text{and} \quad & \exp(2\pi i p \cdot V_i) \ketAkin{p}_L,
\end{eqnarray}
for $i = 1, 2, 3$, and only invariant combinations of right and left movers survive. 
The $\text{E}_8 \times \text{E}_8'$ gauge group of the heterotic string breaks to 
$\text{E}_6 \times \text{U}(1)^2 \times \text{E}_8'$. The remaining 168 roots of the broken 
$\text{E}_8 \times \text{E}_8'$ become matter states 
\begin{center}
$3 \times \boldsymbol{27}$, $3 \times \boldsymbol{\overline{27}}$ and $6 \times \boldsymbol{1}$.
\end{center}
In the \textbf{twisted sector} the mass formula for the left movers changes (see 
eq.\ (\ref{eq:mass_formula_twisted_sector})), because of the change in the zero point energy and because of the shifted 
root lattice $p_\textmd{sh} = p + V_i$
\begin{eqnarray}
\frac{(p + V_i)^2}{2} + \tilde{N} - \frac{3}{4} & = & 0.
\end{eqnarray}
Each constructing element of the space group corresponds to two different mass 
formulas: one with (moded) excitations for the left movers and one without 
excitations. We give an example 
\begin{eqnarray}
(\theta_2, e_3) \Rightarrow 
\begin{cases}
  (p + V_2)^2 = \frac{3}{2} & \quad \text{for } \tilde{N} = 0 \\
  (p + V_2)^2 = \frac{1}{2} & \quad \text{for } \tilde{N} = \frac{1}{2},
\end{cases}
\end{eqnarray}
where $p$ is an element of the $\text{E}_8 \times \text{E}_8'$ lattice. Here, the torus shift 
$e_3$ of the constructing element does not play a role for the mass formulas. 
In the presence of Wilson lines this will change. 

Taking a closer look at the mass formulas, one can see that the equation for 
$\tilde{N} = 0$ allows a wide range of choices for the root vector $p$, leading to 
quite large representations. Compared to this the equation for $\tilde{N} = \frac{1}{2}$ 
is much more restrictive and will mostly lead to singlets. 

\begin{table}[t!]
\center\begin{tabular}{|c|l|l|c|}
\hline
Case    & Shifts & Gauge Group &  Gen.\\
\hline\hline
1 &
\parbox{5cm}{
\vspace{0.1cm}
$\left(\frac{1}{2}, -\frac{1}{2}, 0^6 \right)\left(0^8 \right)$ \\[0.3\baselineskip]
$\left(0, \frac{1}{2}, -\frac{1}{2}, 0^5 \right)\left(0^8 \right)$ \\[-0.3\baselineskip]
} & 
$\text{E}_6 \times \text{U}(1)^2 \times \text{E}_8'$ & 
48\\ \hline
2 & 
\parbox{5cm}{
\vspace{0.1cm}
$\left(\frac{1}{2}, -\frac{1}{2}, 0^6 \right)\left(0^8 \right)$ \\[0.3\baselineskip]
$\left(0, \frac{1}{2}, -\frac{1}{2}, 0^4, 1\right)\left(1, 0^7 \right)$ \\[-0.3\baselineskip]
} &
$\text{E}_6 \times \text{U}(1)^2 \times \text{SO}(16)'$ & 
16\\ \hline
3 & 
\parbox{5cm}{
\vspace{0.1cm}
$\left(\frac{1}{2}^2, 0^6 \right)\left(0^8 \right)$ \\[0.3\baselineskip]
$\left(\frac{5}{4}, \frac{1}{4}^7 \right)\left(\frac{1}{2}, \frac{1}{2}, 0^6\right)$ \\[-0.3\baselineskip]
} &
$\text{SU}(8)\times \text{U}(1) \times \text{E}_7' \times \text{SU}(2)'$ & 
16\\ \hline
4 & 
\parbox{5cm}{
\vspace{0.1cm}
$\left(\frac{1}{2}^2, 0^5, 1 \right)\left(1, 0^7 \right)$ \\[0.3\baselineskip]
$\left(0, \frac{1}{2}, -\frac{1}{2}, 0^5 \right)\left(-\frac{1}{2}, \frac{1}{2}^3, 1, 0^3 \right)$ \\[-0.3\baselineskip]
} &
$\text{E}_6 \times \text{U}(1)^2 \times \text{SO}(8)'^2$ & 
0\\ \hline
5 & 
\parbox{5cm}{
\vspace{0.1cm}
$\left(\frac{1}{2}, -\frac{1}{2}, -1, 0^5\right)\left(1, 0^7 \right)$ \\[0.3\baselineskip]
$\left(\frac{5}{4}, \frac{1}{4}^7 \right)\left(\frac{1}{2}, \frac{1}{2}, 0^6 \right)$  \\[-0.3\baselineskip]
} &
$\text{SU}(8)\times \text{U}(1) \times \text{SO}(12)' \times \text{SU}(2)'^2$ & 
0\\ \hline
\end{tabular}
\caption{Classification of ${\mathbb Z}_2\times {\mathbb Z}_2$ orbifolds without background fields.}
\label{tab:inequivalentshifts}
\end{table}

As described in section \ref{sec:spectrum} the right movers also become twisted. 
As in the untwisted case, right and left movers now have to be sorted 
with respect to their eigenvalues under all shifts. It is important to 
notice that all left movers (of the untwisted sector and of the twisted 
sectors) find a right moving partner to form invariant states. Since the 
states of the twisted sectors are half-hypermultiplets of 
$\mathcal{N} = 2$ supersymmetry, CP conjugation is needed to form complete 
$\mathcal{N} = 1$ chiral multiplets. The eigenvalue of the chirality is 
defined as the first entry of the right moving $\text{SO}(8)$ spinor 
(Ramond state). We choose to count states with negative chirality and combine 
them with their CP partners to get complete multiplets. A CP partner is equal 
to the original state except for a multiplication with $-1$ in the 
$\text{E}_8 \times \text{E}_8'$ root lattice of the left mover and in the 
$\textmd{SO}(8)$ lattice of the right mover. Therefore also the eigenvalues 
of the left and right mover are the same except for a multiplication with 
$-1$. Using this in case of the standard embedding the matter content of 
the twisted sector is
\begin{center}
$48 \times \boldsymbol{27}$ and $240 \times \boldsymbol{1}$, 
\end{center}
where five singlets and one $\boldsymbol{27}$ live on every fixed
torus. Since the untwisted sector has a net number of zero families, the 
standard embedding leads to a chiral spectrum with a net number of 48 families.  

We have classified all {${\mathbb Z}_2\times {\mathbb Z}_2$ orbifold models without
background fields. It turns out that there are only 
five inequivalent combinations of shifts. We present the result in 
table \ref{tab:inequivalentshifts}. The first model (standard
embedding) has already been presented in \cite{Font:1988mk}. 
 
\begin{figure}[b!]
\centerline{\epsfig{figure=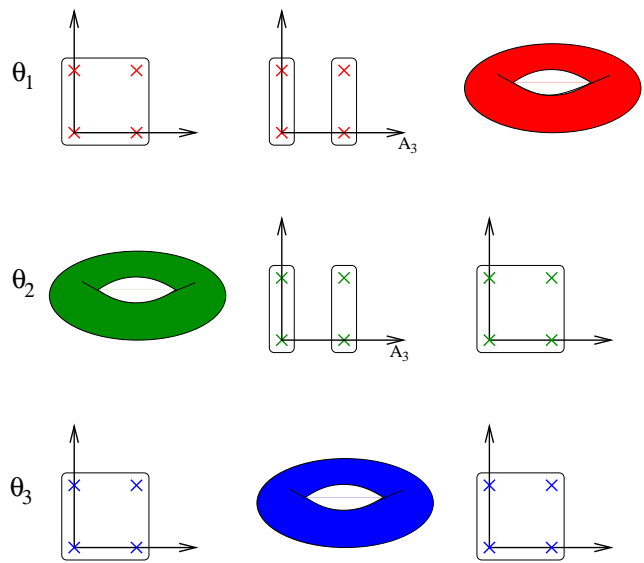,angle=270,scale=.4}}
\caption{Wilson line in $e_3$ direction lifts the degeneracy of the fixed points as 
indicated by the boxes. }
\label{fig:deg_of_fixedpoint}
\end{figure}

\subsection{Adding Wilson Lines}

Wilson lines are needed to get interesting gauge groups and to reduce the
number of families. As explained in section \ref{sec:review_of_orbifolds}, Wilson lines are the embedding 
of the torus shifts into the gauge degrees of freedom. In the untwisted 
sector, states with left movers that are invariant under their action survive, i.e.
\begin{eqnarray}
p \cdot A_i & \in & {\mathbb Z}, \nonumber
\end{eqnarray}
and the other states are projected out. Therefore, they break the 
gauge group. Additionally, Wilson lines control the number of families in 
the twisted sectors. This is due to the fact that Wilson lines can distinguish 
between different fixed points by changing the mass formulas. For example, without 
Wilson lines the constructing elements $(\theta_2, 0)$ and $(\theta_2, e_3)$ 
lead to the same mass formulas. This changes now:
\begin{eqnarray}
(\theta_2, 0) & \Rightarrow &
\begin{cases}
  (p + V_2)^2 = \frac{3}{2} & \quad \text{for }  \tilde{N} = 0 \\
  (p + V_2)^2 = \frac{1}{2} & \quad \text{for }  \tilde{N} = \frac{1}{2}
\end{cases} \\
(\theta_2, e_3) & \Rightarrow & 
\begin{cases}
  (p + V_2 + A_3)^2 = \frac{3}{2} & \quad \text{for }  \tilde{N} = 0 \\
  (p + V_2 + A_3)^2 = \frac{1}{2} & \quad \text{for }  \tilde{N} = \frac{1}{2}
\end{cases}
\end{eqnarray} 
We illustrate the lifting of the degeneracy of the fixed points in figure \ref{fig:deg_of_fixedpoint}.

The $\tilde{N} = \frac{1}{2}$ mass equation in the $(\theta_2, e_3)$ case is 
too restrictive to give any other representations but singlets. But by a clever 
choice of Wilson lines, the 
$\tilde{N} = 0$ mass equation for the same fixed point still allows both: either to have a representation of a family 
or several smaller ones. Hence, Wilson lines reduce the number of families.

A second way in which Wilson lines control the number of families appears
only in presence of fixed tori. A Wilson line that corresponds to a 
direction in a fixed torus acts like an additional projector. 
This is due to the fact that one has to project onto all elements of $Z_g$, 
which is the set of space group elements that commute with the constructing 
element $g$ (section \ref{sec:spectrum}). We give an example. Suppose that 
$g = (\theta_3, e_1)$ is the constructing element. Then the set of commuting 
space group elements $Z_g$ consists of several elements, e.g. the constructing 
element $g$ itself and $(\theta_2, e_3)$
\begin{eqnarray}
\left[(\theta_3, e_1), (\theta_2, e_3)\right] = 0\text{ .}
\end{eqnarray}
Thus one has to calculate the eigenvalues with respect to all elements 
of $Z_g$ for the left and right movers. We show how to calculate the 
eigenvalues for the commuting element $(\theta_2, e_3)$
\begin{eqnarray}
\exp\left(2\pi i p_\textmd{sh} \cdot (V_2 + A_3)\right) \ketAkin{p_\textmd{sh}}_L \quad & \text{and} & \quad
\exp\left(2\pi i q_\textmd{tw} \cdot v_2\right) \ketAkin{q_\textmd{tw}}_R
\end{eqnarray}
where $p_\textmd{sh} = p + V_3 + A_1$ and $q_\textmd{tw} = q + v_3$ correspond to the 
constructing element $(\theta_3, e_1)$. Only invariant combinations of right and left 
movers survive the projection. It is important to notice that due to these additional projections in the presence of Wilson 
lines, not all left movers find a right moving partner 
to form invariant states.

\subsection {SO(10) Model with Three Families}
\label{sec:so10_model}

\begin{figure}[h!]
\centerline{\epsfig{figure=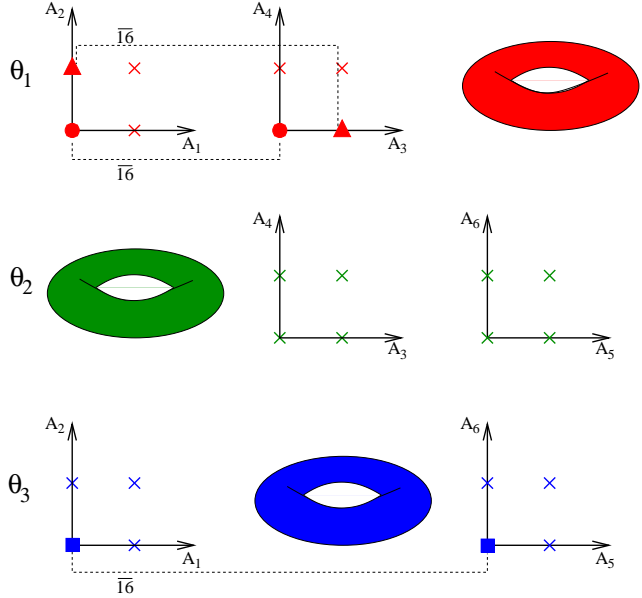,angle=270,scale=.4}}
\caption{Localization of the families for the toy $\text{SO}(10)$ model. }
\label{fig:so10_3families}
\end{figure}
Wilson lines are therefore a promising tool to construct interesting models. 
As a toy model, we present an $\text{SO}(10)$ model with three families. We use standard 
embedding together with the six Wilson lines

\vspace{0.5cm}
{
\renewcommand{\baselinestretch}{1.5}
\normalsize
\begin{tabular}{ll}
$A_1 = (0^7, 1)(1, 0^7)$, & $A_4 = (0^8)(1, 0^3, \frac{1}{2}, -\frac{1}{2}, \frac{1}{2}, \frac{1}{2})$, \\
$A_2 = (0^8)(0, 1, \left(\frac{1}{2}\right)^4, 0^2)$, & $A_5 = (0^8)(0, \frac{1}{2}, -1, \frac{1}{2}, -\frac{1}{2}, 0, \frac{1}{2},0)$, \\
$A_3 = (0^8)(0^2, \frac{1}{2}, -\frac{1}{2}, -\frac{1}{2}, \frac{1}{2}, -1, 0 )$, & $A_6 = (0^8)(1, 0^3, \frac{1}{2}, -\frac{1}{2}, \frac{1}{2}, \frac{1}{2})$.\\
\end{tabular}
}
\vspace{0.5cm}

The first half of $A_1$ breaks $\text{E}_6 \times \text{U}(1)^2$ to $\text{SO}(10) \times \text{U}(1)^3$. 
The other Wilson lines do not break this any further. The hidden $\text{E}_8'$ breaks 
to $\text{U}(1)'^8$. The matter content is 
\begin{center}\begin{tabular}{ll}
untwisted sector: & $12 \times \boldsymbol{1}$ and $6 \times \boldsymbol{10}$\\
twisted sector:   & $174 \times \boldsymbol{1}$, $3 \times \boldsymbol{\overline{16}}$ and $5 \times \boldsymbol{10}$.
\end{tabular}\end{center}
The $\boldsymbol{\overline{16}}$ of $\text{SO}(10)$ counts as a family, thus we have a three family 
toy model\footnote{The existence of three family models in this context
seems to be in apparent contradiction to the results obtained in
\cite{Donagi:2004ht}. This discrepancy can be explained by the fact 
that in \cite{Donagi:2004ht} background fields (Wilson lines) had not
been included.}
of $\text{SO}(10)$. Their localization is illustrated in figure \ref{fig:so10_3families}. 
The two families of the $\theta_1$ sector live on parallel 5-branes and the third family of the $\theta_3$ 
sector lives on an orthogonal brane. Interestingly, not every twisted sector leads to a family. Matter in non-trivial representations under $\text{SO}(10)$ is listed in table \ref{tab:so10}.

\begin{table}[h!]
\begin{center}
\begin{tabular}{c||c|c}
sector & \# of $\boldsymbol{10}$s & \# of  $\boldsymbol{\overline{16}}$s\\
\hline \hline 
untwisted & 6 & 0 \\
\hline
$(\theta_1 , 0)$ & 1 & 1 \\
$(\theta_1 , e_2 + e_3)$  & 1 & 1 \\
\hline
$(\theta_2 , 0)$ &  1 & 0 \\
$(\theta_2 , e_4 + e_6)$ & 1 & 0 \\
\hline
$(\theta_3 , 0)$ & 1 & 1 \\
\hline 
sum & 11 & 3 
\end{tabular}
\end{center}
\caption{$\text{SO}(10)$ model. Matter states in non-trivial representations.}
\label{tab:so10}
\end{table}

\section{Exploiting the Geometric Picture}
\label{sec:exploiting_the_geometric_picture}

One advantage of orbifold compactifications is that they provide a
clear geometric picture. String theory predicts whether certain fields
are constrained to live on a lower dimensional brane or can propagate
through the bulk in a very simple way: Twisted sector states are
constrained to the corresponding fixed plane, whereas untwisted fields
propagate in ten space-time dimensions. In particular, the gauge fields
are always bulk fields in heterotic models. Matter fields can come
from untwisted as well as twisted sectors and hence can be bulk as
well as brane fields. In the following we are going
to exploit the geometric picture for our ${\mathbb Z}_2 \times
{\mathbb Z}_2$ example further. 

\subsection{Localization of Charged Matter}

Here, we discuss the
localization of charged matter in a setting where we zoom in on one of
the compact two-tori. 
Physically this would correspond to a situation in which two of
the extra dimensions are larger than the other four. We should stress,
however, that we will not discuss the size of the extra dimensions here
but merely want to give a detailed geometric picture of our example.

To this end let us zoom in on the {\bf first torus}. First, we
restrict our discussion 
to the three families transforming in the
$\boldsymbol{\overline{16}}$ of SO(10). The families
appear in the $\theta_1$ and $\theta_3$
twisted sector and are localized
at fixed 
points in the first torus:
There is one family at the $\left(
\theta_1 , 0\right)$ fixed point and one at the $\left(
\theta_3 , 0\right)$ fixed point. In the zoomed in picture these 
families are both localized at the origin in the
first torus lattice. A third family lives at the  $\left(\theta_1 , e_2 +
e_3\right)$ fixed point which corresponds to the point $e_2/2$ in the
first torus. The distribution of the families within the first torus
is shown in figure \ref{fig:zoom1}.

For the discussion of Yukawa couplings (see next section) it is also
important where Higgs fields appear in the compact geometry. In our
toy model we do have several fields transforming in the
$\boldsymbol{10}$ of SO(10), i.e.\ many candidates for a standard
model Higgs field. The localization of these $\boldsymbol{10}$s within
the first torus is as follows: In the bulk there are eight fields, six
from the untwisted sector and two from the $\theta_2$ twisted sector,
since the first torus is invariant under $\theta_2$. Further, there
are two $\boldsymbol{10}$s at the origin: one from the $\theta_1$
twisted sector and one from the $\theta_3$ twisted sector. Another
$\boldsymbol{10}$ from the $\theta_1$ twisted sector sits at the point
$e_2/2$.  
\begin{figure}
\begin{center}
\centerline{\epsfig{figure= 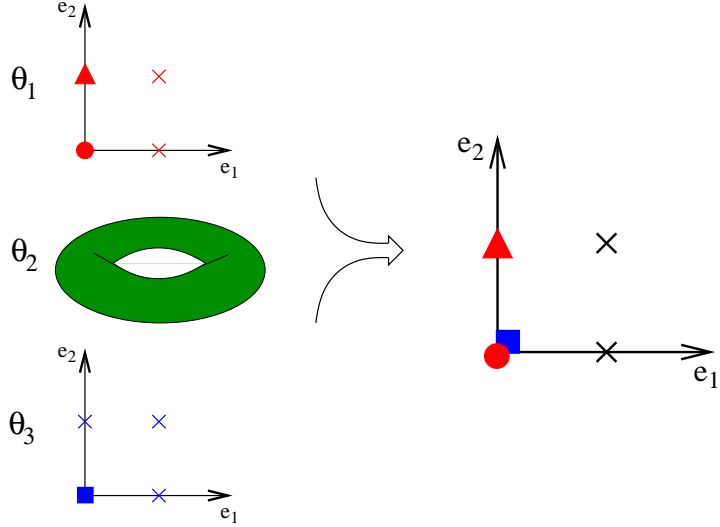,scale=0.7}}
\end{center}
\caption{In this picture we focus on the localization of families
  within the first torus. The left hand side shows the three twisted
  sectors separately, whereas on the right hand side they are merged into
  one representation. For the sake of clarity we do not show the 11
  Higgs candidates. Their localization is given in the
  text.}\label{fig:zoom1} 
\end{figure}

If we zoom in on the {\bf second torus}, the family from the $\theta_3$
twisted sector lives in the bulk. Out of the families from the
$\theta_1$ twisted 
sector, one lives at the origin of the second torus and one at the
fixed point $e_3/2$. The family localization is summarized in figure
\ref{fig:zoom2}. 
Furthermore, there are seven $\boldsymbol{10}$s in the bulk (six from
the untwisted and one from the $\theta_3$ twisted sector). 
From the $\theta_1$ twisted sector one gets one $\boldsymbol{10}$ at
the origin and one $\boldsymbol{10}$ at $e_3/2$.
One $\boldsymbol{10}$ from the $\theta_2$ twisted sector is localized at
the origin and one at $e_4/2$.  
\begin{figure}
\begin{center}
\centerline{\epsfig{figure= 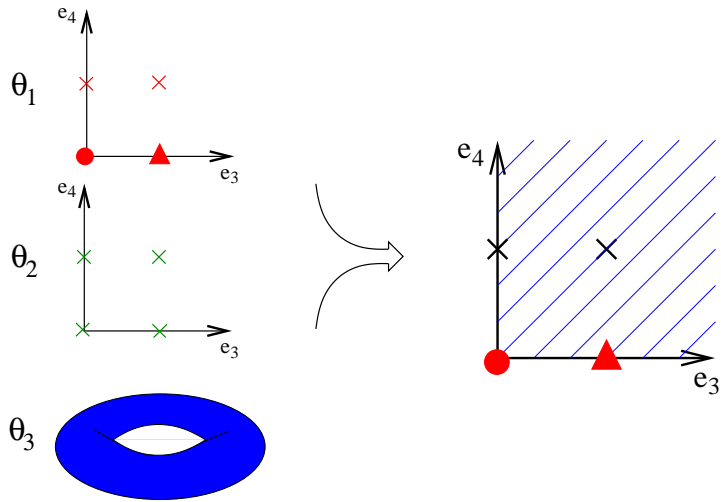,scale=0.7}}
\end{center}
\caption{The family localization within the second torus is shown. The
  diagonal lines symbolize one family living in the bulk. (For more
  explanations, see figure \ref{fig:zoom1}.)}\label{fig:zoom2}
\end{figure}

Finally, zooming in on the {\bf third torus} provides a setup where two
families (from the $\theta_1$ twisted sector) live in the bulk of the
third torus, whereas the family from the $\theta_3$ twisted sector is
localized at the origin. 
The  $\boldsymbol{10}$s are distributed as follows: eight in the bulk
(from untwisted and $\theta_1$ twisted sectors), two at the origin
(from the $\theta_2$ and $\theta_3$ twisted sector), and one at $e_6/2$
(from the $\theta_2$ twisted sector). Figure \ref{fig:zoom3} shows how
the families are localized within the third torus.
\begin{figure}
\begin{center}
\centerline{\epsfig{figure= 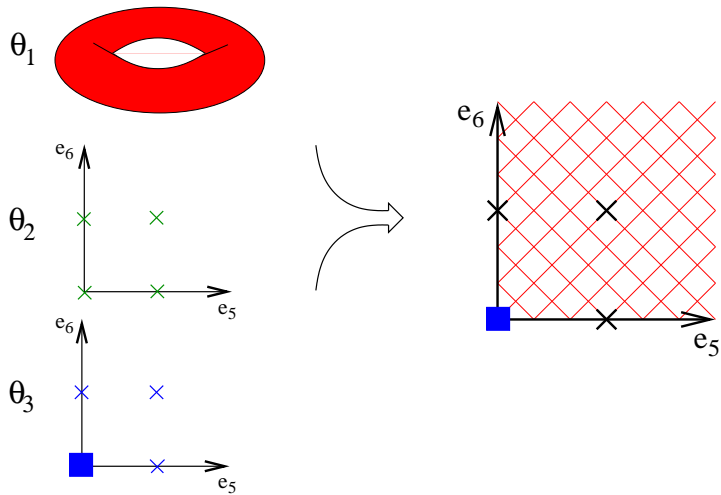,scale=0.7}}
\end{center}
\caption{The family localization within the third torus is shown. The
  diagonal lines symbolize two families living in the bulk. (For more
  explanations, see figure \ref{fig:zoom1}.)}\label{fig:zoom3}
\end{figure}

\subsection{Gauge Group Geography}

In this subsection we are going to provide a picture of the local
physics at the fixed points. Such a description was used in
\cite{Gmeiner:2002es} to develop the concept of fixed point equivalent
models. Fixed point equivalent models yield the same spectrum at a
given fixed point. Since they are usually chosen to have a simpler
structure (no Wilson lines), they are very helpful for answering
questions concerning the local physics at a  fixed point. 
The picture we will present is also useful in order to make 
contact with the so called orbifold 
GUTs. Orbifold GUTs are field theoretic descriptions where one
compactifies a higher dimensional field theory on an orbifold. The
transformation properties of the fields under the orbifold group are
usually given by assigning parities to the fields by hand. In
addition, localized matter is also 
added by hand. (For a review see \cite{review}.) In string theory,
all these data are dictated e.g.\ by 
modular invariance which automatically ensures anomaly free
theories in the higher dimensions as well as in four space-time
dimensions. 

Since the orbifold fixed planes in our model are five-branes, it is
natural to discuss a six dimensional orbifold GUT
picture.\footnote{Any other dimension below or equal ten is also
  possible. This is a special property of ${\mathbb Z}_2 \times
  {\mathbb Z}_2$ since all the radii can be chosen
  freely here.} If we  
choose for example the first torus to represent the extra dimensions
of a 6d orbifold GUT, we would first compactify the heterotic string on
a $T^4/{\mathbb Z}_2$ where the ${\mathbb Z}_2$ twist is given by
$\theta_2$. The resulting spectrum is the bulk spectrum of the
orbifold GUT. In a second step, this six dimensional theory is
compactified on a $T^2/{\mathbb Z}_2$ where the ${\mathbb Z}_2$ is
generated by $\theta_1$. Due to the presence of Wilson lines we have
different projection conditions on different fixed
points. In \cite{Asaka:2001eh,Asaka:2002my} such a situation is
viewed as a 
$T^2/{\mathbb Z}_2 \times {\mathbb Z}_2 \times {\mathbb Z}_2$
orbifold. Since we know how the space group elements containing
$\theta_1$ act on the string states, all the parities are given by our
initial choice of the heterotic orbifold. In addition, we also know
which fields to localize on the fixed points from our analysis of the
twisted sectors.

Just in order to show, that it is not so difficult to find more three 
generation models in ${\mathbb Z}_2 \times {\mathbb Z}_2$ orbifolds we
are going to illustrate the above discussion at the example of a three
generation $\text{SU}(5)$ model. Let us first briefly present this model
leaving out many details (which will be given elsewhere).  The
orbifold shift is again standard embedded, and there are
five Wilson lines along the first five directions in $T^2 \times T^2
\times T^2$:

\vspace{0.5cm}

{
\renewcommand{\baselinestretch}{1.5}
\normalsize
\begin{tabular}{ll}
$A_1 =$ $\left(0^6, \frac{1}{2}, \frac{1}{2}\right)$
  $\left(1,\frac{1}{2},\frac{1}{2},0^5\right)$, &  $A_2 =$ 
$\left(\frac{1}{2},\frac{1}{2},\frac{1}{2}, 0^3, \frac{1}{2},
  0\right)$ $\left( 1, 
  -\frac{1}{2}, 0,\frac{3}{2},\frac{1}{2},\frac{1}{2},1,0\right)$, \\ 
$A_3 =$ $\left(0^8\right)$ $\left(1,1,-1, 0, \frac{1}{2},
  -\frac{1}{2},\frac{1}{2},-\frac{3}{2}\right)$, & 
$A_4 =$ $\left(0^6, \frac{1}{2},\frac{1}{2}\right)$ $\left(\frac{1}{2},
  0^2,-\frac{1}{2}, 0^3,1\right)$, \\ 
$A_5 =$ $\left(0^8\right)$
  $\left(1,1,-1,0,\frac{1}{2},-\frac{1}{2},\frac{1}{2},-\frac{3}{2}\right)$.   
\end{tabular}
}
\vspace{0.5cm}

The four
dimensional gauge group is $\text{SU}(5)$ times $\text{U}(1)$ factors. One family
contains a $\boldsymbol{5}$ and a $\boldsymbol{\overline{10}}$. In the
untwisted 
spectrum there are three $\boldsymbol{5}$s and three
$\boldsymbol{\overline{5}}$s giving a net number of zero families. The
three families arise from various twisted sectors. The relevant matter
is listed in table \ref{tab:su5twi}.
\begin{table}
\begin{center}
\begin{tabular}{c||c|c|c|c}
twist sector & \# of {\bf 5}s &\# of  $\boldsymbol{\overline{5}}$s &
\# of {\bf 10}s & \# of $\boldsymbol{\overline{10}}$s\\
\hline \hline 
$(\theta_1 , 0)$ & 2 & 1 & 0 & 1 \\
$(\theta_1 , e_1)$  & 2 & 0 & 0 & 0 \\
$(\theta_1 , e_3 + e_4)$ & 2 & 0 & 0 & 0 \\
\hline
$(\theta_2 ,0)$ &  0 & 1 & 0 & 0\\
$(\theta_2 , e_3 + e_5)$ & 0 & 1 & 0 & 0\\
$(\theta_2, e_6)$ & 0& 1 & 0 & 0\\
$(\theta_2 , e_3 + e_5 + e_6)$  & 0 & 1 & 0 & 0\\
\hline
$(\theta_3 , 0)$ & 0 &1 & 0 & 1 \\
$(\theta_3 , e_1)$ & 2 & 0 & 0 & 0\\
$(\theta_3 , e_6)$ & 0 & 1 & 0 & 1\\
$(\theta_3 , e_1 + e_6)$ & 2 & 0 & 0 &0  \\
\hline 
sum & 10 & 7 & 0 & 3 
\end{tabular}
\end{center}
\caption{$\text{SU}(5)$ model. Twisted sector states in non-trivial representations.}
\label{tab:su5twi}
\end{table}
Further details of this model will be discussed in a forthcoming
publication. 

Here, we restrict our attention to the interpretation as
a six dimensional field theory orbifold. We choose the first torus to
be the one playing the role of the extra dimensions in the field
theory orbifold. Let us first focus on the pattern of gauge symmetry
breaking. In the bulk of that orbifold we have the gauge group
SO(12)$\times$SU(2)$\times$SU(2). This comes from applying the
projection conditions arising from $\theta_2$ and the Wilson lines
$A_3, A_4 , A_5$ on the E$_8 \times$E$_8$ gauge group. The remaining
orbifold elements relate
the value of the gauge field at a point of the first torus to its
value at the image point. For a fixed point it imposes a projection
condition reducing the size of the gauge group. For example imposing
invariance under $(\theta_1 , 0)$ \footnote{Since $\theta_3 =
  \theta_1\theta_2$ invariance under $(\theta_3,0)$ is ensured.} reduces
SO(12)$\times$SU(2)$\times$SU(2) to SU(6)$\times$SU(2). At the fixed
point  $e_1 /2$ we have to impose invariance under $(\theta_1 ,
e_1)$. In the first E$_8$, the first Wilson line and the fourth Wilson
line are the same. Hence on the bulk gauge group, $(\theta_1,
e_1)$ acts in the same way as $(\theta_1, 0)$, and the bulk symmetry
is broken to the same SU(6)$\times$SU(2). At the fixed point  
$e_2 /2$, however, the bulk symmetry is reduced to SO(10) (by imposing
invariance under $(\theta_1, e_2)$). The same happens at $(e_1 +
e_2)/2$. The situation is illustrated in figure \ref{fig:su5GUT}.
%%%%%%%%%%%%%%%%%%%%%%%%%%%%%%%%%%%%
\begin{figure}[h!]
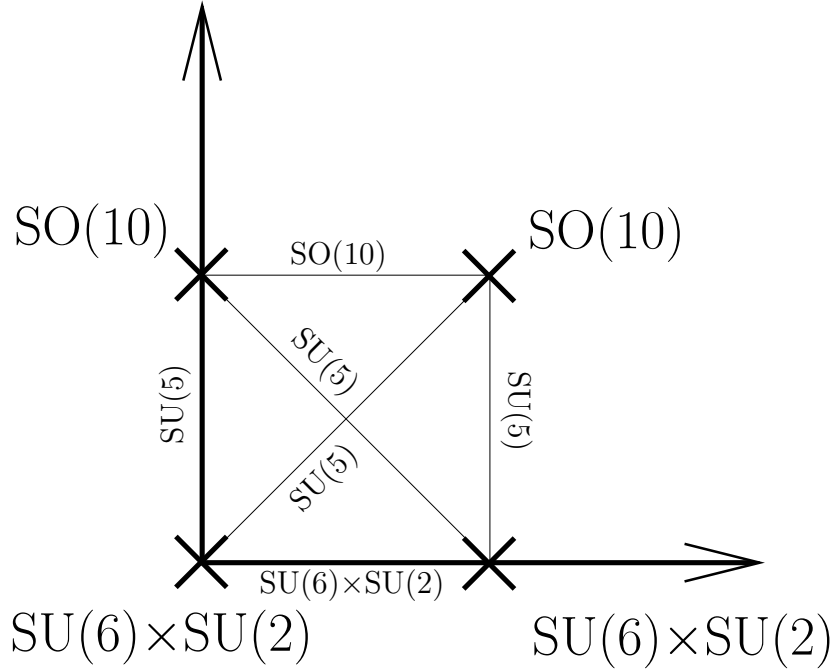
  
\begin{center}
\input gaugegroup_su5.pstex_t
\end{center}
\caption{6d field theory orbifold picture of gauge symmetry breaking
  pattern in the SU(5) model. In the bulk there is an
  SO(12)$\times$SU(2)$\times$SU(2) gauge group. Gauge groups written
  at lines connecting two fixed points are the ones surviving both of
  the corresponding projection conditions.}
\label{fig:su5GUT}
\end{figure}
%%%%%%%%%%%%%%%%%%%%%%%%%%%%%%%%%%%%
Massless gauge fields in four dimensions arise from 6d gauge field
configurations not depending on the extra coordinates. This is
possible only if the gauge field lies in the overlap of the gauge
groups surviving all projection conditions. In our case this leads to
an SU(5) symmetry in four dimensions. The matter from the $\theta_2$
twisted sectors lives in the bulk of the torus, whereas the matter from
the other twisted sectors is localized at the corresponding fixed
points. 

As a second example we want to discuss a model with Pati--Salam gauge
group $\text{SU}(4)\times \text{SU}(2)\times\text{SU}(2)$. (We
suppress the hidden 
sector gauge group and U(1)
factors. The rank of the gauge group is 
never reduced in our models.) Again, this is obtained from
the standard embedding and five additional Wilson lines:

\vspace{0.5cm}
{
\renewcommand{\baselinestretch}{1.5}
\normalsize
\begin{tabular}{ll}
$A_1 =$ $(0^4, \frac{1}{2},\frac{1}{2}, 0^2)$
  $(\frac{1}{2},\frac{1}{2},1,0^5)$, & $A_2 =$ $(0^6, \frac{1}{2} 
,\frac{1}{2})$ $(0^3, \frac{1}{2} ,\frac{1}{2}, 1, 0^2)$,\\ $A_3 =$
  $(0^8)$ $(\frac{1}{2},-\frac{1}{2},1, 
0^3,\frac{1}{2},\frac{1}{2})$,& $A_4 =$ $(0^8)$ $(0^2, -1,
  \frac{1}{2},-\frac{1}{2},0,\frac{1}{2},-\frac{1}{2})$,\\ $A_5 
=$ $(0^8)$ $(\frac{1}{2},-\frac{1}{2},1,0^3 ,\frac{1}{2},\frac{1}{2})$. 
\end{tabular}
}
\vspace{0.5cm}

As generations we count the
$(\boldsymbol{4}, \boldsymbol{2},\boldsymbol{1})$ representation of
the Pati--Salam group. We focus our discussion only on this
representation, leaving out matter transforming
differently. (Equivalently we could count $(\boldsymbol{\overline{4}},
\boldsymbol{1},\boldsymbol{2})$ since these representations come always
together  in the considered model.) There is one generation at the
$(\theta_1, 0)$ fixed point, one generation at the $(\theta_3,
0)$ fixed point and one generation at the $(\theta_3 , e_6)$ fixed
point. 

For the interpretation as a 6d orbifold we take again the first torus
as the one with the extra field theory dimensions. The Wilson lines in
the second and third tori have entries only in the (hidden sector)
second E$_8$. So, the gauge group in the bulk is E$_7 \times$SU(2) in
the observable sector. Here, it appears that the reduced gauge group at
different fixed points is the same (e.g.\ E$_6$), which however can be
embedded differently into the bulk gauge group. Therefore, in figure
\ref{fig:pati-salam} gauge groups written at lines connecting the fixed points 
can be smaller than the gauge groups at the fixed points. In this model
all three generations are located at the origin of the torus.
%%%%%%%%%%%%%%%%%%%%%%%%%%%%%%%%%%%%
\begin{figure}[h!]
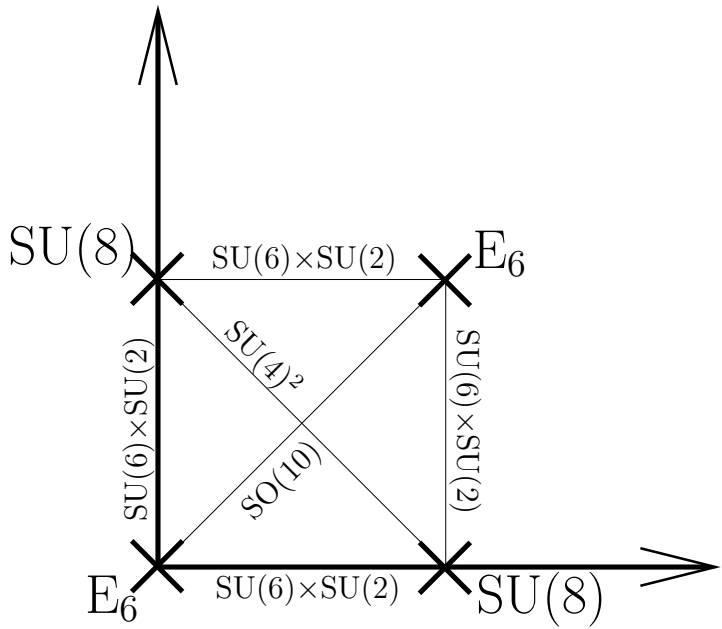
  
\begin{center}
\input gaugegroup_patisalam.pstex_t
\end{center}
\caption{6d field theory orbifold picture of gauge symmetry breaking
  pattern in the model with Pati--Salam gauge group. In the bulk there
  is an E$_7 \times$SU(2). The groups written at the lines are the
  common overlap of the reduced gauge groups appearing at the fixed
  points connected by the line.}
\label{fig:pati-salam}
\end{figure}
%%%%%%%%%%%%%%%%%%%%%%%%%%%%%%%%%%%%

We have seen that orbifold GUTs are incorporated into heterotic
orbifolds in a very natural way. Consistency is guaranteed due to the
underlying consistent string theory. A similar discussion can be found
in \cite{Kobayashi:2004ud} where a ${\mathbb Z}_6 = {\mathbb Z}_3
\times {\mathbb Z}_2$ model with unbroken Pati-Salam group is
presented. These authors derived a five dimensional field theory
orbifold from a heterotic model. This, of course, can also be done in
a straightforward way in our model. Indeed, in the ${\mathbb Z}_2
\times {\mathbb Z}_2$ model one has the maximal freedom in choosing
the radii. 
Finally, we would like to point out that knowledge about the gauge
group geography is also relevant for conceptual questions like local
anomaly cancellation
\cite{GrootNibbelink:2002wv,GrootNibbelink:2002qp,Lee:2003mc,Gmeiner:2002es,GrootNibbelink:2003gb,Nibbelink:2003rc}.

\section{Towards Realistic Models}
\label{sec:Towards_realistic_models}

So far we have discussed just a few simple models to 
illustrate properties of the heterotic brane world.
The conditions outlined in section \ref{sec:review_of_orbifolds} have been 
meanwhile incorporated 
in computer programs that allow 
the efficient and fast construction of many new models, 
in fact, many more as we are able to classify. 
Of course, we are primarily interested in the construction 
of realistic models that contain the standard 
model of particle physics and we have to develop a 
strategy to select interesting models 
guided by phenomenological requirements. 

Especially in the framework of the ${\mathbb Z}_N \times {\mathbb
  Z}_M$ picture  
we expect a multitude of promising models.
In this paper, however, we shall not focus on explicit 
discussion of such models and refer the reader to an 
upcoming publication \cite{future}. Instead we shall discuss the 
properties of the heterotic brane world models at the 
qualitative level to point out which phenomenological 
questions can be addressed successfully within this 
picture.

\subsection{Phenomenological Restrictions}
The questions we hope to answer in this scheme will be concerned with

\medskip

{\bf Gauge coupling unification}. With this we mean 
an explanation of the values of the 
gauge couplings of $\text{SU}(3)\times \text{SU}(2)\times \text{U}(1)$
from the  
string coupling constant. Note that this 
does not necessarily require the notion of a grand unified 
group in $d=4$. Still  we would like 
to understand the correct value of $\sin^2 \Theta_W$ as 
well as possible threshold corrections. For 
an earlier discussion see \cite{Nilles:1995kb}.

\medskip

{\bf Yukawa coupling unification}. As in the case of the 
gauge couplings we would like to link the values 
of Yukawa couplings to the unified string coupling. We 
should try to see whether a given model allows for 
a parameterization of the correct pattern of quark and lepton 
 masses as well as mixing angles. In particular, 
we would like to identify the explanation of a suppression
 mechanism for some of the couplings as the 
origin for the hierarchy of quark and lepton masses. 
This analysis will include a calculation of 
world sheet instanton contribution to the effective 
superpotential. In the past such a discussion has been 
given in 
\cite{Casas:1988vk,Kobayashi:1990mc,Casas:1990hi,Casas:1991ac}.
For more recent discussion see 
\cite{Lebedev:2001qg,Kobayashi:2003vi,Ko:2004ic}. 
We expect a geometrical explanation of such a pattern 
in the heterotic brane world picture.

\medskip

{\bf Baryon- and lepton-number violation}. Typically we
 have to worry about the stability of the proton. 
Can we hope for a (discrete) symmetry like R-parity to 
avoid problems with proton decay?

\medskip

{\bf Gauge hierarchy problem}. Why is the weak 
scale so small compared to the string scale? We will
 assume the presence of ${\cal N}=1$ supersymmetry in $d=4$. 
But even then we have to solve the doublet-triplet 
 splitting problem (hopefully through split multiplets 
as in \cite{Ibanez:1987sn})
 and the so-called  $\mu$ problem, with 
 $\mu$ being the mass parameter in the Higgs superpotential. 
Is there a connection with axions as a 
 solution to the strong CP-problem  \cite{Kim:1983dt}? 
Supersymmetry has to 
be broken at a scale small compared to the 
 string scale. Can we create a hidden sector responsible for 
that breakdown 
\cite{Derendinger:1985kk,Dine:rz,Derendinger:1985cv,Nilles:2004zg}, 
as e.g.\ the $\text{E}_8'$ 
 sector of the heterotic $\text{E}_8 \times \text{E}_8'$ theory?

\medskip

 {\bf Gauge symmetry breakdown}. This question 
might be relevant at several stages; e.g. the 
 breakdown of a grand unified group, or the breakdown 
of weak interaction $\text{SU}(2) \times \text{U}(1)$. Often, 
 the rank reduction of the underlying gauge group needs 
a specific mechanism 
\cite{Ibanez:1987xa,Font:1988tp}, 
which very often 
 boils down to the breakdown of additional $\text{U}(1)$ gauge bosons.

 There are, of course, more detailed questions to be asked for
 realistic model building (absence of flavour changing neutral
 currents, origin of  
 CP-violation just to name a few) that acquire the knowledge of very 
 specific properties of the models under consideration. In our discussion 
 here we shall, however, first concentrate on the more qualitative issues 
 quoted above in the framework of a geometrical picture.

 \subsection{Properties of Heterotic Orbifolds}
 So let us now inspect key properties of heterotic brane world models 
 in view of the phenomenological applications. 

\medskip

  {\bf Gauge group}. It is a subgroup of $\text{E}_8 \times \text{E}_8'$ or $\text{SO}(32)$. 
  We concentrate here on $\text{E}_8 \times \text{E}_8'$ as this theory is phenomenologically 
  preferable. These groups lead to chiral fermions in $d=10$ but not 
  in $d=4$. Thus we need a subgroup that allows for parity violation in $d=4$.
  This could be a grand unified theory like the $\text{SO}(10)$ and $\text{SU}(5)$ toy 
  examples given in the earlier sections, smaller groups 
like $\text{SU}(4) \times \text{SU}(2) \,{\mbox{and}}\, \text{SU}(3)^3$ 
  or just the standard model gauge group $\text{SU}(3) \times \text{SU}(2) \times \text{U}(1)$. 
  The latter would be preferable since it allows the presence 
of split multiplets 
\cite{Ibanez:1987sn} 
and in addition we do not need 
  to incorporate the Higgs multiplets for the spontaneous 
breakdown of the grand unified gauge group. 
  In fact, it turns out to be practically impossible to 
obtain the necessary representations in the framework 
  of realistic $\text{SO}(10)$ and $\text{SU}(5)$ models \cite{Lewellen:1989qe}. 
In the intermediate cases (like Pati-Salam group or 
  $\text{SU}(3)^3$ trinification \cite{Choi:2003ag}) 
such Higgs fields could be 
present as they originate from a (split) 27-dimensional 
  representation of an underlying $\text{E}_6$. At this point we should mention
a weakness of the construction explained 
  in this paper. It does not allow the reduction of the rank 
of the gauge group. This is the price we have 
  to pay for the simplicity of the construction. 
Thus if we start with 
  $\text{E}_8 \supset \text{SU}(3) \times \text{SU}(2) \times \text{U}(1) $ we will
 have four additional $\text{U}(1)$ factors. Rank reduction needs 
  more input, as e.g. the implementation of continuous Wilson lines 
\cite{Ibanez:1987xa} 
or the consideration of so-called 
  degenerate orbifolds 
 \cite{Font:1988tp}.
We shall not discuss this here in detail.
  So let us now consider a theory with standard model gauge group 
in $d=4$. Although this is not a 
  bona fide GUT model in $d=4$ it might inherit a lot of the 
successful properties 
  of e.g.\ the $\text{SO}(10)$ or $\text{E}_6$ theory. A family of quarks and 
leptons is in the 
  16-dimensional representation of $\text{SO}(10)$ which 
also contains an R-symmetry that 
  forbids fast proton decay by dimension 4 operators in 
the supersymmetric framework. These 
  are remnants of the underlying grand unified group in $d > 4$ such as 
  $\text{SU}(5) \subset \text{SO}(10) \subset \text{E}_6 \subset \text{E}_7 \subset \text{E}_8$. 
Even better, some of the problems
  of GUTs (such as the doublet-triplet splitting problem) 
are solved because Higgs bosons 
  (as well as gauge bosons) appear in incomplete (split) 
multiplets. Many of the phenomenological 
  properties of the $d=4$ theory will depend on  
the degree to which it remembers its grand 
  unified origin in $d > 5$. This includes the value 
of $\sin^2 \Theta_W$, the question of proton 
  stability and the unification of Yukawa couplings. 
To understand these remnants of higher 
  dimensional grand unification it is extremely useful 
to examine the geography of gauge 
  group realizations like those shown in figure \ref{fig:su5GUT} and \ref{fig:pati-salam} for 
the example under consideration. In 
  connection with the knowledge of the localization of 
matter and Higgs fields we can read off 
  allowed and forbidden couplings, as Yukawa interactions
and B, L violating operators. At the 
  various locations, gauge groups are typically enhanced with respect to 
  $\text{SU}(3) \times \text{SU}(2)\times \text{U}(1)$ and this might forbid 
unwanted operators and couplings. 
  As we shall see in 
\cite{future} 
these properties will prove 
to be useful for realistic model building.       

\medskip

{\bf Spectrum of matter and Higgs fields}. One would 
aim at the constructions of models 
with 3 net families of quarks and leptons. This is certainly 
true for models with standard model 
gauge group. At an intermediate step, however, one might also 
consider models with a grand 
unified group and a different number of matter families. The 
reason for this comes from the 
fact that in our approach with quantized Wilson lines a gauge 
symmetry breakdown is usually 
accompanied by a change in the number of families. If we 
consider, for example, our 
$\text{SO}(10)$ model from section \ref{sec:zn_x_zm} (with 3 families) and consider 
another Wilson line to break 
the gauge symmetry we would then obtain the wrong number of families. 
In that sense some 
other $\text{SO}(10)$ model with a different number of families would 
represent the underlying 
grand unified picture of a 3 family standard model. 
Another point to stress is the 
possible presence of anti-generations. In fact, models  
with just 3 generations usually 
give only limited flexibility to accommodate a realistic 
pattern of Yukawa couplings. 
Therefore, it might be advisable to search for models 
with $n > 3$ families and $n-3$ 
anti-families as well. Very often, the number of 
families is connected to the geometrical 
properties of the model. In the early construction of 
the $\mathbb{Z}_3$ orbifold families 
could be obtained in the untwisted sector and the 
number 3 found its explanation in the 
number of complex compactified dimensions 
\cite{Ibanez:1986tp}. 
In 
other cases a factor 3 appeared 
because of the appearance of 3 twisted sectors 
\cite{Kim:en}. 
Our $\text{SO}(10)$ model in section \ref{sec:so10_model} has 
3 twisted sectors with 2, 1, 0 families respectively 
(${\mathbb Z}_2 \times {\mathbb Z}_2$ models always have a 
zero net number of families in the untwisted sector). 
The locations of the families are 
important for a detailed discussion of the pattern of 
Yukawa couplings. For this we also 
need the location of the candidate Higgs fields 
to break $\text{SU}(2) \times \text{U}(1)$: i.e.\ Higgs 
doublets. In our $\text{SO}(10)$ toy model we have several 10-dimensional 
representations that 
could provide such doublets, but in addition they contain the 
partner $\text{SU}(3)$-triplets and 
we will eventually have a doublet-triplet splitting problem. 
Therefore, we should aim 
for models where only a smaller non-abelian gauge group 
like $\text{SU}(3) \times \text{SU}(2)$ is 
realized in $d=4$ which allows for split multiplets. 
Very often, the models contain 
other exotic representations. One should carefully 
investigate in what sense such 
exotic states could be a signal of string theory in 
the low-energy spectrum. Such 
fields, charged under $\text{SU}(3) \times \text{SU}(2) \times \text{U}(1)$ 
might be highly relevant 
for the evolution of the gauge coupling constants. The 
tree-level gauge coupling 
constant (in particular $\sin^2 \Theta_W$) are strongly 
dependent on the way how 
$\text{U}(1)$-hypercharge is embedded in the (usually) several 
$\text{U}(1)$'s other than hypercharge. The 
appearance of the  $\text{U}(1)$'s and the singlets is an 
artifact of the simplicity of our 
construction and we have to rely on other methods to 
reduce the rank of the gauge group, 
e.g. continuous Wilson lines 
\cite{Ibanez:1987xa}. 
From the low-energy
 point of view, such a mechanism 
corresponds to singlet fields receiving non-vanishing 
vacuum expectation values that break the gauge 
group 
\cite{Font:1988tp}. 
Many of the singlets have flat directions 
in the effective potential and are 
therefore genuine string moduli \cite{Font:1988mm}.

\medskip

{\bf Supersymmetry}. Throughout this discussion 
we assume ${\cal N}=1$ supersymmetry in 
$d=4$. This should help in solving the hierarchy problem. 
But, as we know, ${\cal N}=1$ supersymmetry 
is not enough. We have to deal with the doublet-triplet 
splitting problem as well 
(here the possible appearance of Higgs triplets). 
In fact, the orbifold picture presented 
here constitutes the only known working mechanism to achieve doublet-triplet 
splitting consistently. But even this is not enough, as we have to deal with 
potential Higgs mass terms in the superpotential: the so-called $\mu$-problem. 
Very often, models contain more than 2 doublets. One has 
then to understand why the 
additional doublets become heavy and 2 remain light. 
In a given model such mass terms 
are typically connected to the vacuum expectation 
values of the (singlet) moduli fields. 
In that sense the value of $\mu$ might be coming 
from a mechanism as discussed in 
\cite{Kim:1983dt} 
or \cite{Giudice:1988yz} 
in the field theory case. This might be 
connected with the axion solution 
of the strong CP-problem. Apart from this we have eventually to set up
schemes for a breakdown of supersymmetry, most probably in the framework
of hidden sector gaugino condensation which naturally might be
connected to the properties of the descendants of the $\text{E}_8^\prime$ 
(for a review see \cite{Nilles:2004zg}).

\medskip

{\bf Global (discrete) symmetries}. Apart from the 
gauge symmetries we usually 
find a large number of global (discrete) symmetries 
that might be relevant for low energy 
phenomenology. Very often they come from the symmetries 
of the orbifold 
${\mathbb Z}_N$ or ${\mathbb Z}_N \times {\mathbb Z}_M$. They might
also originate  
as discrete subgroups of 
underlying gauge symmetries. Such symmetries might 
be important for the flavour structure 
of the model, patterns of quark mass matrices and 
potential appearance of rare processes. In 
particular this concerns the stability of the proton. 
The string models generically  do 
violate Baryon- and Lepton-number and we need additional 
symmetries to avoid too fast 
proton decay. The usual R-parity of the minimal supersymmetric 
standard model 
(or a variant thereof) needs to be present in realistic models. 
A way to incorporate 
this symmetry might be to profit from the underlying $\text{SO}(10)$ 
structure of the 
specific model under considerations. It allows the standard 
Yukawa couplings and assures 
the stability of the proton. Many of the successful models 
incorporate the robust $\text{SO}(10)$ 
relic and sufficient proton stability could be achieved in a 
way that is not strongly 
dependent on the specific geometrical structure of the orbifold.

\section{Outlook}
\label{sec:outlook}

It is now straightforward to search for realistic models of particle
physics. The conditions outlined in  
section 2 can be implemented in computer programs that allow the
construction of many 3 family models,  
in fact so many that we have to apply further selection criteria. We
find it appropriate here to  
restrict the search for models with standard model gauge group $\text{SU}(3)
\times \text{SU}(2) \times \text{U}(1)$ in  
$d=4$ to avoid further problems with spontaneous gauge symmetry
breakdown and to obtain doublet-triplet  
splitting. We require three quark-lepton families but stress that
models with a non-vanishing number of  
anti-families could be preferable in view of the Yukawa coupling
structure. A further selection criterion  
should be the presence of an underlying GUT structure, as e.g.\ $\text{SO}(10)$ or $\text{E}_6$, at some
level in the higher dimensional  
picture. This should ensure that a family of quarks and leptons
transforms effectively as a 16-dimensional  
spinor of $\text{SO}(10)$, although only $\text{SU}(3) \times \text{SU}(2) \times \text{U}(1)$ is
realized in $d=4$: gauge bosons  
and Higgs bosons come in split multiplets of the GUT group but the matter
families do not.  

Such an underlying $\text{SO}(10)$ structure is useful for realistic model
building. It will  
\begin{itemize}
\item give the correct value of $\sin^2 \Theta_w$ at the
  large scale,
 \item allow for a satisfactory implementation of (Majorana)
 neutrino masses,
 \item provide the R-symmetry needed to forbid proton decay
 via dimension 4 operators.
 \end{itemize}
 
 Implementing this successful properties of grand unification in
 models with only  
 $\text{SU}(3) \times \text{SU}(2) \times \text{U}(1)$ gauge group should be the key to
 realistic model building.  
 In this respect, the consideration of ${\mathbb Z}_N \times
 {\mathbb Z}_M$ orbifolds 
 seems to be most promising. 
 
 Unfortunately, the mechanism of quantized Wilson lines does not allow
 rank reduction of the  
 gauge group. We therefore have to face the presence of various $\text{U}(1)$
 gauge groups. Usually,  
 the identification of the hypercharge $\text{U}(1)$ can be quite cumbersome,
 but an underlying GUT  
 structure will simplify this task. In any case the charges of all the
 representations with respect  
 to all of these $\text{U}(1)$'s have to be determined. This will then allow
 the determination of allowed  
 couplings in the superpotential as well as the determination of the
 (singlet) moduli fields.  
 Rank reduction could occur through the vacuum expectation values of
 such fields \cite{Font:1988tp}. It might  
 also give an explicit realization of the blowing-up procedure in
 orbifold compactification in a  
 low-energy effective field theory approximation. Within a full string
 theory mechanism, rank  
 reduction can be achieved through continuous Wilson lines
 \cite{Ibanez:1987xa}. The 
 inclusion of this  
 mechanism within the context of realistic model building should be
 pursued \cite{future}. 
 
 As we said, a key geometrical property of the heterotic orbifold scheme is the potential appearance of fixed tori or fixed points. In this paper we have illustrated the geometrical picture with
 the help of some toy models.  
 New realistic models have been identified and will be presented
 in detail in a future  
 publication. Related work has recently appeared in
 \cite{Kobayashi:2004ud} in the 
 framework of a ${\mathbb Z}_6$-model with  
 $\text{SU}(4) \times \text{SU}(2) \times \text{SU}(2)$ gauge group. In the framework of the fermionic formulation of the heterotic string theory a $\mathbb{Z}_2\times \mathbb{Z}_2$ twist has been discussed in \cite{Faraggi:2004rq}.
 
 Ultimately, one would like to incorporate the M-theory picture of
 Ho\v{r}ava and Witten \cite{Horava:1995qa} into our  
 framework, as it provides a geometrical interpretation of the
 supersymmetry breakdown in the  
 hidden sector \cite{Horava:1996vs,Nilles:1997cm} as well. The theory,
 however, is not yet well enough understood. More work  
 along the lines of \cite{Gorbatov:2001pw,Conrad:2000tk} is needed.

\bigskip

\noindent {\bf Acknowledgments} 
 
\noindent 
It is a pleasure to thank Mark Hillenbach and Martin
Walter for discussions. Especially, we would like to thank David Grellscheid for his assistance. This work is supported by the European Commission RTN programs
\mbox{HPRN-CT-2000-00131}, 00148 and 00152.


\begin{thebibliography}{42}
%use Latex-EU style from spires

%\cite{Narain:1985jj}
\bibitem{Narain:1985jj}
K.~S.~Narain,
%``New Heterotic String Theories In Uncompactified Dimensions < 10,''
Phys.\ Lett.\ B {\bf 169} (1986) 41.
%%CITATION = PHLTA,B169,41;%%

%\cite{Candelas:en}
\bibitem{Candelas:en}
P.~Candelas, G.~T.~Horowitz, A.~Strominger and E.~Witten,
%``Vacuum Configurations For Superstrings,''
Nucl.\ Phys.\ B {\bf 258} (1985) 46.
%%CITATION = NUPHA,B258,46;%%


%\cite{Dixon:jw}
\bibitem{Dixon:jw}
L.~J.~Dixon, J.~A.~Harvey, C.~Vafa and E.~Witten,
%``Strings On Orbifolds,''
Nucl.\ Phys.\ B {\bf 261} (1985) 678.
%%CITATION = NUPHA,B261,678;%%


%\cite{Dixon:1986jc}
\bibitem{Dixon:1986jc}
L.~J.~Dixon, J.~A.~Harvey, C.~Vafa and E.~Witten,
%``Strings On Orbifolds. 2,''
Nucl.\ Phys.\ B {\bf 274} (1986) 285.
%%CITATION = NUPHA,B274,285;%%


%\cite{Ibanez:1986tp}
\bibitem{Ibanez:1986tp}
L.~E.~Ib\'{a}\~{n}ez, H.~P.~Nilles and F.~Quevedo,
%``Orbifolds And Wilson Lines,''
Phys.\ Lett.\ B {\bf 187} (1987) 25.
%%CITATION = PHLTA,B187,25;%%

%\cite{Ibanez:1987sn}
\bibitem{Ibanez:1987sn}
L.~E.~Ib\'{a}\~{n}ez, J.~E.~Kim, H.~P.~Nilles and F.~Quevedo,
%``Orbifold Compactifications With Three Families Of SU(3) X SU(2) X U(1)**N,''
Phys.\ Lett.\ B {\bf 191} (1987) 282.
%%CITATION = PHLTA,B191,282;%%

%\cite{Kawamura:2000ir}
\bibitem{Kawamura:2000ir}
Y.~Kawamura,
%``Split multiplets, coupling unification and extra dimension,''
Prog.\ Theor.\ Phys.\  {\bf 105} (2001) 691
[arXiv:hep-ph/0012352].
%%CITATION = HEP-PH 0012352;%%

%\cite{Kawamura:2000ev}
\bibitem{Kawamura:2000ev}
Y.~Kawamura,
%``Triplet-doublet splitting, proton stability and extra dimension,''
Prog.\ Theor.\ Phys.\  {\bf 105} (2001) 999
[arXiv:hep-ph/0012125].
%%CITATION = HEP-PH 0012125;%%

%\cite{Altarelli:2001qj}
\bibitem{Altarelli:2001qj}
G.~Altarelli and F.~Feruglio,
%``SU(5) grand unification in extra dimensions and proton decay,''
Phys.\ Lett.\ B {\bf 511} (2001) 257
[arXiv:hep-ph/0102301].
%%CITATION = HEP-PH 0102301;%%


%\cite{Hall:2001pg}
\bibitem{Hall:2001pg}
L.~J.~Hall and Y.~Nomura,
%``Gauge unification in higher dimensions,''
Phys.\ Rev.\ D {\bf 64} (2001) 055003
[arXiv:hep-ph/0103125].
%%CITATION = HEP-PH 0103125;%%


%\cite{Kawamoto:2001wm}
\bibitem{Kawamoto:2001wm}
T.~Kawamoto and Y.~Kawamura,
%``Symmetry reduction, gauge transformation and orbifold,''
arXiv:hep-ph/0106163.
%%CITATION = HEP-PH 0106163;%%

%\cite{Hebecker:2001wq}
\bibitem{Hebecker:2001wq}
A.~Hebecker and J.~March-Russell,
%``A minimal S(1)/(Z(2) x Z'(2)) orbifold GUT,''
Nucl.\ Phys.\ B {\bf 613} (2001) 3
[arXiv:hep-ph/0106166].
%%CITATION = HEP-PH 0106166;%%


\bibitem{Asaka:2001eh}
T.~Asaka, W.~Buchm\"uller and L.~Covi,
%``Gauge unification in six dimensions,''
Phys.\ Lett.\ B {\bf 523} (2001) 199
[arXiv:hep-ph/0108021].
%%CITATION = HEP-PH 0108021;%%


%\cite{Hamidi:1986vh}
\bibitem{Hamidi:1986vh}
S.~Hamidi and C.~Vafa,
%``Interactions On Orbifolds,''
Nucl.\ Phys.\ B {\bf 279} (1987) 465.
%%CITATION = NUPHA,B279,465;%%

%\cite{Dixon:1986qv}
\bibitem{Dixon:1986qv}
L.~J.~Dixon, D.~Friedan, E.~J.~Martinec and S.~H.~Shenker,
%``The Conformal Field Theory Of Orbifolds,''
Nucl.\ Phys.\ B {\bf 282} (1987) 13.
%%CITATION = NUPHA,B282,13;%%

%\cite{Lauer:1989ax}
\bibitem{Lauer:1989ax}
J.~Lauer, J.~Mas and H.~P.~Nilles,
%``Duality And The Role Of Nonperturbative Effects On The World Sheet,''
Phys.\ Lett.\ B {\bf 226} (1989) 251.
%%CITATION = PHLTA,B226,251;%%

%\cite{Lauer:1990tm}
\bibitem{Lauer:1990tm}
J.~Lauer, J.~Mas and H.~P.~Nilles,
%``Twisted Sector Representations Of Discrete Background Symmetries For
%Two-Dimensional Orbifolds,''
Nucl.\ Phys.\ B {\bf 351} (1991) 353.
%%CITATION = NUPHA,B351,353;%%

%\cite{Burwick:1990tu}
\bibitem{Burwick:1990tu}
T.~T.~Burwick, R.~K.~Kaiser and H.~F.~M\"uller,
%``General Yukawa Couplings Of Strings On Z(N) Orbifolds,''
Nucl.\ Phys.\ B {\bf 355} (1991) 689.
%%CITATION = NUPHA,B355,689;%%

%\cite{Stieberger:1992bj}
\bibitem{Stieberger:1992bj}
S.~Stieberger, D.~Jungnickel, J.~Lauer and M.~Spalinski,
%``Yukawa couplings for bosonic Z(N) orbifolds: Their moduli and twisted sector
%dependence,''
Mod.\ Phys.\ Lett.\ A {\bf 7} (1992) 3059
[arXiv:hep-th/9204037].
%%CITATION = HEP-TH 9204037;%%

%\cite{Erler:1992gt}
\bibitem{Erler:1992gt}
J.~Erler, D.~Jungnickel, M.~Spalinski and S.~Stieberger,
%``Higher twisted sector couplings of Z(N) orbifolds,''
Nucl.\ Phys.\ B {\bf 397} (1993) 379
[arXiv:hep-th/9207049].
%%CITATION = HEP-TH 9207049;%%


%\cite{Kaplunovsky:1987rp}
\bibitem{Kaplunovsky:1987rp}
V.~S.~Kaplunovsky,
%``One Loop Threshold Effects In String Unification,''
Nucl.\ Phys.\ B {\bf 307} (1988) 145
[Erratum-ibid.\ B {\bf 382} (1992) 436]
[arXiv:hep-th/9205068].
%%CITATION = HEP-TH 9205068;%%


%\cite{Dixon:1990pc}
\bibitem{Dixon:1990pc}
L.~J.~Dixon, V.~Kaplunovsky and J.~Louis,
%``Moduli Dependence Of String Loop Corrections To Gauge Coupling Constants,''
Nucl.\ Phys.\ B {\bf 355} (1991) 649.
%%CITATION = NUPHA,B355,649;%%

%\cite{Mayr:1993kn}
\bibitem{Mayr:1993kn}
P.~Mayr, H.~P.~Nilles and S.~Stieberger,
%``String unification and threshold corrections,''
Phys.\ Lett.\ B {\bf 317} (1993) 53
[arXiv:hep-th/9307171].
%%CITATION = HEP-TH 9307171;%%

%\cite{Mayr:1995rx}
\bibitem{Mayr:1995rx}
P.~Mayr and S.~Stieberger,
%``Moduli dependence of one loop gauge couplings in (0,2) compactifications,''
Phys.\ Lett.\ B {\bf 355} (1995) 107
[arXiv:hep-th/9504129].
%%CITATION = HEP-TH 9504129;%%

%\cite{Font:1988tp}
\bibitem{Font:1988tp}
A.~Font, L.~E.~Ib\'{a}\~{n}ez, H.~P.~Nilles and F.~Quevedo,
%``Degenerate Orbifolds,''
Nucl.\ Phys.\ B {\bf 307} (1988) 109
[Erratum-ibid.\ B {\bf 310} (1988) 764].
%%CITATION = NUPHA,B307,109;%%

%\cite{Font:1988mm}
\bibitem{Font:1988mm}
A.~Font, L.~E.~Ib\'{a}\~{n}ez, H.~P.~Nilles and F.~Quevedo,
%``Yukawa Couplings In Degenerate Orbifolds: Towards A Realistic SU(3) X SU(2) X
%U(1) Superstring,''
Phys.\ Lett.\  {\bf 210B} (1988) 101
[Erratum-ibid.\ B {\bf 213} (1988) 564].
%%CITATION = PHLTA,210B,101;%%

%\cite{Kim:en}
\bibitem{Kim:en}
H.~B.~Kim and J.~E.~Kim,
%``An Orbifold Compactification With Three Families From Twisted Sectors,''
Phys.\ Lett.\ B {\bf 300} (1993) 343
[arXiv:hep-ph/9212311].
%%CITATION = HEP-PH 9212311;%%

%\cite{Font:1988mk}
\bibitem{Font:1988mk}
A.~Font, L.~E.~Ib\'{a}\~{n}ez and F.~Quevedo,
%``Z(N) X Z(M) Orbifolds And Discrete Torsion,''
Phys.\ Lett.\ B {\bf 217} (1989) 272.
%%CITATION = PHLTA,B217,272;%%

%\cite{Bailin:pf}
\bibitem{Bailin:pf}
D.~Bailin, A.~Love and S.~Thomas,
%``Gauge Symmetry Breaking In Orbifold Compactified Superstring Theories,''
Nucl.\ Phys.\ B {\bf 288} (1987) 431.
%%CITATION = NUPHA,B288,431;%%

%\cite{Bailin:1986pd}
\bibitem{Bailin:1986pd}
D.~Bailin, A.~Love and S.~Thomas,
%``Fermion Generations In Orbifold Compactified Superstring Theories,''
Phys.\ Lett.\ B {\bf 188} (1987) 193.
%%CITATION = PHLTA,B188,193;%%

%\cite{Bailin:1987xm}
\bibitem{Bailin:1987xm}
D.~Bailin, A.~Love and S.~Thomas,
%``A Three Generation Orbifold Compactified Superstring Model With Realistic
%Gauge Group,''
Phys.\ Lett.\ B {\bf 194} (1987) 385.
%%CITATION = PHLTA,B194,385;%%

%\cite{Bailin:1987dm}
\bibitem{Bailin:1987dm}
D.~Bailin, A.~Love and S.~Thomas,
%``Generalized Gso Projections For Z Orbifold Heterotic String Compactifications
%With 'Wilson Lines',''
Mod.\ Phys.\ Lett.\ A {\bf 3} (1988) 167.
%%CITATION = MPLAE,A3,167;%%

%\cite{Casas:1987us}
\bibitem{Casas:1987us}
J.~A.~Casas, E.~K.~Katehou and C.~Mu\~{n}oz,
%``U(1) Charges In Orbifolds: Anomaly Cancellation And Phenomenological
%Consequences,''
Nucl.\ Phys.\ B {\bf 317} (1989) 171.
%%CITATION = NUPHA,B317,171;%%

%\cite{Casas:1988se}
\bibitem{Casas:1988se}
J.~A.~Casas and C.~Mu\~{n}oz,
%``Three Generation SU(3) X SU(2) X U(1)-Y X U(1) Orbifold Models Through
%Fayet-Iliopoulos Terms,''
Phys.\ Lett.\ B {\bf 209} (1988) 214.
%%CITATION = PHLTA,B209,214;%%

%\cite{Casas:1988hb}
\bibitem{Casas:1988hb}
J.~A.~Casas and C.~Mu\~{n}oz,
%``Three Generation SU(3) X SU(2) X U(1)-Y Models From Orbifolds,''
Phys.\ Lett.\ B {\bf 214} (1988) 63.
%%CITATION = PHLTA,B214,63;%%

%\cite{Katsuki:1988ku}
\bibitem{Katsuki:1988ku}
Y.~Katsuki, Y.~Kawamura, T.~Kobayashi and N.~Ohtsubo,
%``Z(7) Orbifold Models,''
Phys.\ Lett.\ B {\bf 212} (1988) 339.
%%CITATION = PHLTA,B212,339;%%

%\cite{Katsuki:1989kd}
\bibitem{Katsuki:1989kd}
Y.~Katsuki, Y.~Kawamura, T.~Kobayashi, N.~Ohtsubo and K.~Tanioka,
%``Gauge Groups Of Z(N) Orbifold Models,''
Prog.\ Theor.\ Phys.\  {\bf 82} (1989) 171.
%%CITATION = PTPKA,82,171;%%

%\cite{Katsuki:1989ra}
\bibitem{Katsuki:1989ra}
Y.~Katsuki, Y.~Kawamura, T.~Kobayashi, N.~Ohtsubo, Y.~Ono and K.~Tanioka,
%``Z(8) And Z(12) Orbifold Models,''
Phys.\ Lett.\ B {\bf 227} (1989) 381.
%%CITATION = PHLTA,B227,381;%%

%\cite{Quevedo:1996sv}
\bibitem{Quevedo:1996sv}
F.~Quevedo,
%``Lectures on superstring phenomenology,''
arXiv:hep-th/9603074.
%%CITATION = HEP-TH 9603074;%%

%\cite{Hwang:2002hg}
\bibitem{Hwang:2002hg}
K.~w.~Hwang and J.~E.~Kim,
%``Orbifolded SU(7) and unification of families,''
Phys.\ Lett.\ B {\bf 540} (2002) 289
[arXiv:hep-ph/0205093].
%%CITATION = HEP-PH 0205093;%%

%\cite{Kim:2003ch}
\bibitem{Kim:2003ch}
J.~E.~Kim,
%``Z(3) orbifold construction of SU(3)**3 GUT with sin**2(Theta(0)(W)) =  3/8,''
Phys.\ Lett.\ B {\bf 564} (2003) 35
[arXiv:hep-th/0301177].
%%CITATION = HEP-TH 0301177;%%

%\cite{Choi:2003pq}
\bibitem{Choi:2003pq}
K.~S.~Choi, K.~Hwang and J.~E.~Kim,
%``Dynkin diagram strategy for orbifolding with Wilson lines,''
Nucl.\ Phys.\ B {\bf 662} (2003) 476
[arXiv:hep-th/0304243].
%%CITATION = HEP-TH 0304243;%%



%\cite{Choi:2003ag}
\bibitem{Choi:2003ag}
K.~S.~Choi and J.~E.~Kim,
%``Three family Z(3) orbifold trinification, MSSM and doublet-triplet  splitting
%problem,''
Phys.\ Lett.\ B {\bf 567} (2003) 87
[arXiv:hep-ph/0305002].
%%CITATION = HEP-PH 0305002;%%


%\cite{Kim:2003hr}
\bibitem{Kim:2003hr}
J.~E.~Kim,
%``SU(3) trits of orbifolded E(8) x E'(8) heterotic string and supersymmetric
%standard model,''
JHEP {\bf 0308} (2003) 010
[arXiv:hep-ph/0308064].
%%CITATION = HEP-PH 0308064;%%

%\cite{Kim:2004pe}
\bibitem{Kim:2004pe}
J.~E.~Kim,
%``Trinification with sin**2(theta(W)) = 3/8 and seesaw neutrino mass,''
Phys.\ Lett.\ B {\bf 591} (2004) 119
[arXiv:hep-ph/0403196].
%%CITATION = HEP-PH 0403196;%%

%\cite{Giedt:2003an}
\bibitem{Giedt:2003an}
J.~Giedt,
%``Z(3) orbifolds of the SO(32) heterotic string: 1 Wilson line  embeddings,''
Nucl.\ Phys.\ B {\bf 671} (2003) 133
[arXiv:hep-th/0301232].
%%CITATION = HEP-TH 0301232;%%

%\cite{Giedt:2004wd}
\bibitem{Giedt:2004wd}
J.~Giedt,
%``Lack of trinification in Z(3) orbifolds of the SO(32) heterotic string,''
arXiv:hep-ph/0402201.
%%CITATION = HEP-PH 0402201;%%

%\cite{Choi:2004vb}
\bibitem{Choi:2004vb}
K.~S.~Choi,
%``Spectrum of heterotic string on orbifold,''
arXiv:hep-th/0405195.
%%CITATION = HEP-TH 0405195;%%

%\cite{Hall:2002ea}
\bibitem{review}
L.~J.~Hall and Y.~Nomura,
%``Grand unification in higher dimensions,''
Annals Phys.\  {\bf 306} (2003) 132
[arXiv:hep-ph/0212134].
%%CITATION = HEP-PH 0212134;%%

%\cite{Kobayashi:2004ud}
\bibitem{Kobayashi:2004ud}
T.~Kobayashi, S.~Raby and R.~J.~Zhang,
%``Constructing 5d orbifold grand unified theories from heterotic strings,''
arXiv:hep-ph/0403065.
%%CITATION = HEP-PH 0403065;%%


%\cite{GrootNibbelink:2002wv}
\bibitem{GrootNibbelink:2002wv}
S.~Groot Nibbelink, H.~P.~Nilles and M.~Olechowski,
%``Spontaneous localization of bulk matter fields,''
Phys.\ Lett.\ B {\bf 536} (2002) 270
[arXiv:hep-th/0203055].
%%CITATION = HEP-TH 0203055;%%

%\cite{GrootNibbelink:2002qp}
\bibitem{GrootNibbelink:2002qp}
S.~Groot Nibbelink, H.~P.~Nilles and M.~Olechowski,
%``Instabilities of bulk fields and anomalies on orbifolds,''
Nucl.\ Phys.\ B {\bf 640} (2002) 171
[arXiv:hep-th/0205012].
%%CITATION = HEP-TH 0205012;%%

%\cite{Lee:2003mc}
\bibitem{Lee:2003mc}
H.~M.~Lee, H.~P.~Nilles and M.~Zucker,
%``Spontaneous localization of bulk fields: The six-dimensional case,''
Nucl.\ Phys.\ B {\bf 680} (2004) 177
[arXiv:hep-th/0309195].

%\cite{Gmeiner:2002es}
\bibitem{Gmeiner:2002es}
F.~Gmeiner, S.~Groot Nibbelink, H.~P.~Nilles, M.~Olechowski and M.~G.~A.~Walter,
%``Localized anomalies in heterotic orbifolds,''
Nucl.\ Phys.\ B {\bf 648} (2003) 35
[arXiv:hep-th/0208146].
%%CITATION = HEP-TH 0208146;%%

%\cite{GrootNibbelink:2003gb}
\bibitem{GrootNibbelink:2003gb}
S.~Groot Nibbelink, H.~P.~Nilles, M.~Olechowski and M.~G.~A.~Walter,
%``Localized tadpoles of anomalous heterotic U(1)'s,''
Nucl.\ Phys.\ B {\bf 665} (2003) 236
[arXiv:hep-th/0303101].
%%CITATION = HEP-TH 0303101;%%

%\cite{Nibbelink:2003rc}
\bibitem{Nibbelink:2003rc}
S.~G.~Nibbelink, M.~Hillenbach, T.~Kobayashi and M.~G.~A.~Walter,
%``Localization of heterotic 
%anomalies on various hyper surfaces of T(6)/Z(4),''
Phys.\ Rev.\ D {\bf 69} (2004) 046001
[arXiv:hep-th/0308076].
%%CITATION = HEP-TH 0308076;%%

%\cite{Uranga:nq}
\bibitem{Uranga:nq}
A.~M.~Uranga,
%``Local Intersecting Brane Worlds,''
Fortsch.\ Phys.\  {\bf 51} (2003) 879.
%%CITATION = FPYKA,51,879;%%

\bibitem{future}
S.~F\"orste, H.~P.~Nilles, P.~K.~S.~Vaudrevange, A.~Wingerter, {\it to appear}

%\cite{Ibanez:1987pj}
\bibitem{Ibanez:1987pj}
L.~E.~Ib\'{a}\~{n}ez, J.~Mas, H.~P.~Nilles and F.~Quevedo,
%``Heterotic Strings In Symmetric And Asymmetric Orbifold Backgrounds,''
Nucl.\ Phys.\ B {\bf 301} (1988) 157.
%%CITATION = NUPHA,B301,157;%%

%\cite{Donagi:2004ht}
\bibitem{Donagi:2004ht}
R.~Donagi and A.~E.~Faraggi,
%``On the number of chiral generations in Z(2) x Z(2) orbifolds,''
arXiv:hep-th/0403272.
%%CITATION = HEP-TH 0403272;%%


%\cite{Faraggi:2004rq}
\bibitem{Faraggi:2004rq}
A.~E.~Faraggi, C.~Kounnas, S.~E.~M.~Nooij and J.~Rizos,
%``Classification of the chiral Z(2) x Z(2) fermionic models in the heterotic
%superstring,''
arXiv:hep-th/0403058.
%%CITATION = HEP-TH 0403058;%%

\bibitem{Asaka:2002my}
T.~Asaka, W.~Buchm\"uller and L.~Covi,
%``Bulk and brane anomalies in six dimensions,''
Nucl.\ Phys.\ B {\bf 648} (2003) 231
[arXiv:hep-ph/0209144].


%\cite{Nilles:1995kb}
\bibitem{Nilles:1995kb}
H.~P.~Nilles and S.~Stieberger,
%``How to Reach the Correct sin~2\theta_W and \alpha_S in String Theory,''
Phys.\ Lett.\ B {\bf 367} (1996) 126
[arXiv:hep-th/9510009].
%%CITATION = HEP-TH 9510009;%%


%\cite{Casas:1988vk}
\bibitem{Casas:1988vk}
J.~A.~Casas and C.~Mu\~{n}oz,
%``Yukawa Couplings In SU(3) X SU(2) X U(1)-Y Orbifold Models,''
Phys.\ Lett.\ B {\bf 212} (1988) 343.
%%CITATION = PHLTA,B212,343;%%

%\cite{Kobayashi:1990mc}
\bibitem{Kobayashi:1990mc}
T.~Kobayashi and N.~Ohtsubo,
%``Yukawa Coupling Condition Of Z(N) Orbifold Models,''
Phys.\ Lett.\ B {\bf 245} (1990) 441.
%%CITATION = PHLTA,B245,441;%%


%\cite{Casas:1990hi}
\bibitem{Casas:1990hi}
J.~A.~Casas, F.~Gomez and C.~Mu\~{n}oz,
%``World Sheet Instanton Contribution To Z(7) Yukawa Couplings,''
Phys.\ Lett.\ B {\bf 251} (1990) 99.
%%CITATION = PHLTA,B251,99;%%


%\cite{Casas:1991ac}
\bibitem{Casas:1991ac}
J.~A.~Casas, F.~Gomez and C.~Mu\~{n}oz,
%``Complete structure of Z(n) Yukawa couplings,''
Int.\ J.\ Mod.\ Phys.\ A {\bf 8} (1993) 455
[arXiv:hep-th/9110060].
%%CITATION = HEP-TH 9110060;%%

%\cite{Lebedev:2001qg}
\bibitem{Lebedev:2001qg}
O.~Lebedev,
%``The CKM phase in heterotic orbifold models,''
Phys.\ Lett.\ B {\bf 521} (2001) 71
[arXiv:hep-th/0108218].
%%CITATION = HEP-TH 0108218;%%


%\cite{Kobayashi:2003vi}
\bibitem{Kobayashi:2003vi}
T.~Kobayashi and O.~Lebedev,
%``Heterotic Yukawa couplings and continuous Wilson lines,''
Phys.\ Lett.\ B {\bf 566} (2003) 164
[arXiv:hep-th/0303009].
%%CITATION = HEP-TH 0303009;%%

%\cite{Ko:2004ic}
\bibitem{Ko:2004ic}
P.~w.~Ko, T.~Kobayashi and J.~h.~Park,
%``Quark masses and mixing angles in heterotic orbifold models,''
arXiv:hep-ph/0406041.
%%CITATION = HEP-PH 0406041;%%


%\cite{Kim:1983dt}
\bibitem{Kim:1983dt}
J.~E.~Kim and H.~P.~Nilles,
%``The Mu Problem And The Strong CP Problem,''
Phys.\ Lett.\ B {\bf 138} (1984) 150.
%%CITATION = PHLTA,B138,150;%%

%\cite{Derendinger:1985kk}
\bibitem{Derendinger:1985kk}
J.~P.~Derendinger, L.~E.~Ib\'{a}\~{n}ez and H.~P.~Nilles,
%``On The Low-Energy D = 4, N=1 Supergravity Theory Extracted From The D = 10,
%N=1 Superstring,''
Phys.\ Lett.\ B {\bf 155} (1985) 65.
%%CITATION = PHLTA,B155,65;%%

%\cite{Dine:rz}
\bibitem{Dine:rz}
M.~Dine, R.~Rohm, N.~Seiberg and E.~Witten,
%``Gluino Condensation In Superstring Models,''
Phys.\ Lett.\ B {\bf 156} (1985) 55.
%%CITATION = PHLTA,B156,55;%%


%\cite{Derendinger:1985cv}
\bibitem{Derendinger:1985cv}
J.~P.~Derendinger, L.~E.~Ib\'{a}\~{n}ez and H.~P.~Nilles,
%``On The Low-Energy Limit Of Superstring Theories,''
Nucl.\ Phys.\ B {\bf 267} (1986) 365.
%%CITATION = NUPHA,B267,365;%%

%\cite{Nilles:2004zg}
\bibitem{Nilles:2004zg}
H.~P.~Nilles,
%``Gaugino condensation and SUSY breakdown,''
arXiv:hep-th/0402022.
%%CITATION = HEP-TH 0402022;%%

%\cite{Ibanez:1987xa}
\bibitem{Ibanez:1987xa}
L.~E.~Ib\'{a}\~{n}ez, H.~P.~Nilles and F.~Quevedo,
%``Reducing The Rank Of The Gauge Group In Orbifold Compactifications Of The
%Heterotic String,''
Phys.\ Lett.\ B {\bf 192} (1987) 332.
%%CITATION = PHLTA,B192,332;%%

%\cite{Lewellen:1989qe}
\bibitem{Lewellen:1989qe}
D.~C.~Lewellen,
%``Embedding Higher Level Kac-Moody Algebras In Heterotic String Models,''
Nucl.\ Phys.\ B {\bf 337} (1990) 61.
%%CITATION = NUPHA,B337,61;%%

%\cite{Giudice:1988yz}
\bibitem{Giudice:1988yz}
G.~F.~Giudice and A.~Masiero,
%``A Natural Solution To The Mu Problem In Supergravity Theories,''
Phys.\ Lett.\ B {\bf 206} (1988) 480.
%%CITATION = PHLTA,B206,480;%%


\bibitem{Horava:1995qa}
P.~Ho\v{r}ava and E.~Witten,
%``Heterotic and type I string dynamics from eleven dimensions,''
Nucl.\ Phys.\ B {\bf 460} (1996) 506
[arXiv:hep-th/9510209].
%%CITATION = HEP-TH 9510209;%%
%
%\cite{Horava:1996vs}
\bibitem{Horava:1996vs}
P.~Ho\v{r}ava,
%``Gluino condensation in strongly coupled heterotic string theory,''
Phys.\ Rev.\ D {\bf 54} (1996) 7561
[arXiv:hep-th/9608019].
%%CITATION = HEP-TH 9608019;%%

\bibitem{Nilles:1997cm}
H.~P.~Nilles, M.~Olechowski and M.~Yamaguchi,
%``Supersymmetry breaking and soft terms in M-theory,''
Phys.\ Lett.\ B {\bf 415} (1997) 24
[arXiv:hep-th/9707143].
%%CITATION = HEP-TH 9707143;%%
%
\bibitem{Conrad:2000tk}
J.~O.~Conrad,
%``On fractional instanton numbers in six dimensional heterotic E(8) x  E(8)
%orbifolds,''
JHEP {\bf 0011} (2000) 022
[arXiv:hep-th/0009251].
%%CITATION = HEP-TH 0009251;%%
\bibitem{Gorbatov:2001pw}
E.~Gorbatov, V.~S.~Kaplunovsky, J.~Sonnenschein, S.~Theisen and
S.~Yankielowicz, 
%``On heterotic orbifolds, M-theory and type I' brane engineering,''
JHEP {\bf 0205} (2002) 015
[arXiv:hep-th/0108135].
%%CITATION = HEP-TH 0108135;%%
%


\end{thebibliography}
\end{document}